\documentclass[conference,compsoc]{IEEEtran}

\usepackage[T1]{fontenc}
\usepackage[utf8]{inputenc}
\usepackage{cite}
\usepackage{amsmath,amssymb}
\usepackage{booktabs}
\usepackage{graphicx}
\usepackage{xcolor}
\usepackage{listings}
\usepackage[skins,breakable,listings]{tcolorbox}
\usepackage{tikz}
\usetikzlibrary{arrows.meta,positioning}
\usepackage{url}
\usepackage[hidelinks]{hyperref}

\newcommand{\tool}{\textsc{WarpGuard}}
\lstdefinestyle{wglisting}{
  basicstyle=\ttfamily\footnotesize,
  columns=fullflexible,
  keepspaces=true,
  showstringspaces=false,
  breaklines=true
}
\newtcblisting{wgcode}{
  listing only,
  listing options={style=wglisting},
  enhanced,
  breakable,
  colback=black!2,
  colframe=black!35,
  boxrule=0.25pt,
  arc=1pt,
  left=0.45em,
  right=0.25em,
  top=0.25em,
  bottom=0.25em,
  boxsep=0.15em,
  before skip=0.45em,
  after skip=0.45em
}

\title{\tool: Protected-Site Control-Flow Integrity for CUDA SASS Binaries}

\author{Igor Santos-Grueiro\\
International University of La Rioja}
\begin{document}

\maketitle

\begin{abstract}
Recent CUDA exploitation work shows that GPU memory bugs can escalate into device-side control-flow corruption, as kernels later consume corrupted return continuations, function pointers, dispatch-table entries, or branch targets. For deployed CUDA binaries, the relevant security boundary is executed NVIDIA SASS, after PTX lowering, inlining, ABI decisions, register allocation, spills, predication, and SIMT execution; source- or PTX-level policies do not capture this boundary.

We present \tool{}, to our knowledge the first protected-site CFI system for CUDA device binaries operating on executed SASS. \tool{} enforces at protected sites: recovered SASS instructions or sequences that consume control-flow state, provide sufficient binary evidence to derive policy, are checked before release, and fail closed on violation. It authenticates backward-edge continuation state for instrumented returns, validates recoverable forward targets per site, and reports fixed-edge, unsupported, profile-excluded, fallback, and no-surface outcomes outside the protected denominator.

On 77 CUDA artifacts, \tool{} classifies 51,621 SASS control-flow sites, including 1,343 returns and 154 supported forward target-set entries, and records 52.2 million dynamic checks. In representative backward- and forward-edge corruption attacks, native execution reaches attacker-selected behavior, detect-only mode records the expected violation, and enforcement fails closed before releasing the invalid protected transfer. Public-code evidence shows that the same SASS consumption patterns occur in real CUDA systems, including runtime dispatch tables, cuFFT callbacks, generated callable tables, and uploaded device-function pointers. \tool{} delivers auditable protected-site CFI for CUDA SASS and separates dynamic-instrumentation enforcement from callback-free SASS timing and patch-cache feasibility.

\end{abstract}

\section{Introduction}

A CUDA memory bug can become a device-side control-flow bug. A later GPU
instruction may consume the corrupted value as a return continuation, a device
function pointer, a dispatch-table entry, or an indirect branch target. The
security event therefore includes both the write that corrupts memory and the
later instruction that releases corrupted state as control flow.

Prior work shows that this is a real CUDA attack surface. Earlier CUDA overflow
studies showed stack, heap, and function-pointer corruption in GPU programs
\cite{di2015cudabuffer,di2016overflow}. Guo et al. demonstrated CUDA code
injection, return-address corruption, and device-side code reuse on modern GPUs
\cite{guo2024gpu}. Roels, Jacobs, and Volckaert showed input-triggered
ROP-style execution entirely inside NVIDIA CUDA device code and measured gadget
expressiveness in SASS binaries \cite{jacobs2025cuda}. These attacks raise the
CFI question inside the GPU: \emph{after memory corruption, which device-side control
transfers should still be allowed?}

Control-Flow Integrity (CFI) is the natural defense family.
It restricts dynamic transfers to policy-approved targets
\cite{abadi2005cfi,burow2017cfisurvey}. CPU CFI also gives a warning: legal
targets can still be abused, and incomplete policies leave useful attack paths
\cite{davi2014stitching,carlini2015control,evans2015controljujutsu}.

For deployed CUDA kernels, the usual CFI boundary is the wrong abstraction.
Production kernels often arrive as fatbinaries, cubins, generated kernels, or
binary libraries. CUDA source is
compiled to PTX and then lowered to native NVIDIA instructions, or SASS
\cite{nvidia_nvcc,nvidia_ptx,nvidia_binary_utils}. The final SASS contains the
control-flow facts: inlining, ABI lowering, register allocation, spills, calls,
returns, predication, and reconvergence. SIMT execution adds lane masks and
warp-uniform or lane-local dynamic targets
\cite{nvidia_programming_guide,shoushtary2024controlflow}.

The central challenge in CUDA binary CFI is deciding which executed SASS
sites have enough evidence to support a sound check. Some transfers are fixed.
Some source calls are inlined and have no
dynamic return surface. Some indirect sites expose recoverable target sets.
Others lack enough binary evidence and must not be silently treated as allowed.
Coverage is therefore not a single whole-binary number. It is a set of
accounted outcomes, and only some of them can support enforcement.

\tool{} enforces CFI only at recovered protected sites and treats every other
outcome as outside the protected denominator. A protected site is a recovered
SASS consumption point whose policy
evidence is sufficient, whose check executes before release, and whose invalid
transfer fails closed. Writable metadata is authenticated under trusted
backend-private state; ordinary device memory does not authorize a transfer.

\tool{} implements this property as a load-time CFI layer for CUDA SASS
binaries. It recovers SASS control-flow sites, binds policy to the loaded image,
authenticates backward-edge continuation state for protected returns, validates
recoverable forward targets per site, and fails closed before releasing invalid
protected transfers.

The evaluation asks whether \tool{} can recover SASS CFI sites, keep the
protected denominator explicit, preserve benign checked transfers, and stop
corrupted control state at protected consumption sites. On 77 CUDA binary
artifacts, \tool{} classifies 51,621 SASS control-flow sites, including 1,343
returns, 9,727 return-terminated gadgets, and 154 supported forward target-set
entries. Runtime counters record 52.2 million dynamic checks.

The security cases use a native/detect/enforce protocol. In protected attack
cases, native execution reaches attacker-selected behavior, detect-only mode
records the expected violation, and enforcement fails closed before releasing
the invalid protected transfer. External public-code cases cover runtime device
tables, cuFFT callback consumers, generated callable tables, uploaded
device-function pointers, and public return consumption. The strongest public
cases include PPL-CUDA-SMC~\cite{repo_ppl_cuda_smc} device dispatch tables and
PNNL SV-Sim/DM-Sim~\cite{repo_sv_sim,repo_dm_sim} uploaded
\texttt{Gate::op} pointers; they show the same SASS consumption shapes outside
artifact fixtures.

WG-NVBit provides broad reference enforcement by instrumenting loaded SASS.
WG-ST and WG-PC provide scoped callback-free evidence: matched static-trampoline
timing and reviewed patch-cache prevention for supported \texttt{sm\_89}
surfaces. We keep these backend denominators separate.

This paper makes three contributions.

\begin{itemize}
  \item \textbf{Protected-site CUDA SASS CFI.} We formulate CUDA binary CFI as
  check-before-release enforcement at recovered SASS consumption sites. To our
  knowledge, this is the first evaluated CUDA device-binary CFI system at
  executed SASS, with explicit accounting for fixed-edge, unsupported, fallback,
  profile-excluded, and no-surface outcomes.
  \item \textbf{Mechanism and implementation.} We build \tool{}, which binds
  policy to recovered SASS, authenticates backward-edge continuation state for
  protected returns, validates recoverable forward targets per site, and keeps
  telemetry outside the authorization path.
  \item \textbf{Evaluation with separated denominators.} We evaluate 77 CUDA
  artifacts, 51,621 classified SASS control-flow sites, 52.2 million dynamic
  checks, representative backward- and forward-edge corrupted-control-state
  attacks, public-code consumption cases, and separated WG-NVBit/WG-ST/WG-PC
  backend evidence.
\end{itemize}

\subsubsection*{Artifact Availability}
The anonymized \tool{} tool and reproduction artifact are available at
\url{https://anonymous.4open.science/r/warpguard-anon/}.

\section{Background and Motivation}

% This section establishes the central boundary used by the rest of the paper:
% CUDA CFI is recovered-site CFI over executed SASS.

\subsection{From Source Structure to SASS Control Flow}

CUDA programs are launched by CPU code, but the control-flow events studied here
happen inside GPU device code. The schematic kernel pattern used throughout the
paper has three source objects: local state, a helper call, and a device
function-pointer dispatch: \texttt{local[idx] = in[tid]},
\texttt{value = helper(local, idx)}, and
\texttt{ops[selector](out, value)}.

The SASS level is where the CFI policy applies. The same source pattern can
produce protected, no-surface, or unsupported outcomes because lowering decides
whether calls survive, where return state lives, and whether indirect targets
are recoverable (Table~\ref{tab:sass-observations}).

The local SASS-boundary audit follows this split: among 125 artifact classifications, 86
have no CFI surface, 22 are backward-only, 4 are forward-only, and 13 expose
both directions. At function granularity, 104 functions have no device-side return,
292 are callsite-only, and 440 contain checked returns
(Table~\ref{tab:app-sass-residuals}).
\subsection{CUDA Control-Flow Attacks Happen at Consumption Sites}

\begin{table}[t!]
\centering
\caption{Source-visible structure does not determine the SASS CFI surface.}
\label{tab:sass-observations}
\footnotesize
\begin{tabular}{@{}p{0.47\linewidth}p{0.46\linewidth}@{}}
\toprule
\textbf{Source/SASS observation} & \textbf{CFI consequence} \\
\midrule
Source helper call is inlined & no backward site \\
Helper becomes \texttt{CALL}/\texttt{RET} & possible protected return \\
Return state stays in register & not writable stack exposure \\
Return state spills/reloads & candidate backward exposure \\
Dispatch table has recoverable targets & protected forward site \\
Indirect target evidence is missing & unsupported, not protected \\
\bottomrule
\end{tabular}
\end{table}

GPU memory corruption can affect correctness, crash kernels, or redirect device
control flow.
Early CUDA overflow studies showed that GPU programs expose stack, heap, and
function-pointer corruption patterns familiar from CPU software
\cite{di2015cudabuffer,di2016overflow}. Guo et al. then showed that modern CUDA
memory bugs can lead to code injection, return-address corruption, and
device-side code reuse \cite{guo2024gpu}. Roels, Jacobs, and Volckaert showed
input-triggered ROP-style execution inside CUDA device code and measured gadget
expressiveness in SASS binaries \cite{jacobs2025cuda}. Other recent GPU work on
ASLR and GPU-to-host memory attacks further weakens the assumption that device
execution is isolated or low-value \cite{zhu2026gpuaslr,roh2026ghost}.

%For CFI, the important point is where these attacks take effect. 
A kernel can
consume corrupted device state as control flow without a host-side hijack. A
bounds bug can overwrite a device function pointer or dispatch-table entry. A
local-memory overwrite can reach spilled continuation state. A data-dependent
indirect branch can receive an attacker-chosen target. In all three cases the
memory bug is only the first stage. The control-flow event happens later, at
the SASS instruction that releases a \texttt{RET}, indirect call, or indirect
branch.
Public CUDA reports often surface as invalid-PC, illegal-instruction, or
illegal-address failures. We only consider them as evidence when corrupted
state reaches a recovered control-flow transfer.

\tool{} does not try to classify every GPU memory bug. It protects the later
SASS instruction that consumes corrupted state as control flow. This explains
why CFI is complementary to GPU memory-safety defenses.
Memory-safety tools try to stop or diagnose the corrupting read or write.
\tool{} checks the consumption point. If a protected transfer is about to use a
return continuation or indirect target outside the recovered policy,
enforcement fails closed before the transfer is released. The goal is to prevent
a memory bug from becoming control-flow release at a protected SASS site.

The policy must therefore be stated and enforced over executed SASS: a buffer
overflow into ordinary data becomes a CFI event only when a SASS instruction
consumes the corrupted value as a target.

\subsubsection*{CUDA/SASS facts}

CUDA device code is compiled to PTX and then to native NVIDIA SASS
\cite{nvidia_ptx,nvidia_nvcc}. SASS contains the CFI-relevant decisions:
inlining, call and return lowering, register allocation, spills, predication,
and exact indirect-transfer operands. NVIDIA GPUs execute SASS in SIMT warps;
predication and active-lane masks mean that a transfer may be inactive for some
lanes, warp-uniform in one execution, and lane-local in another. Check
granularity is therefore part of the policy. Local memory also matters: it is
per-thread in the programming model, but backed by device memory and used for
spills.

CFI restricts dynamic transfers to policy-approved targets
\cite{abadi2005cfi,burow2017cfisurvey}. CPU binary-CFI work shows why
source-free CFI matters \cite{zhang2013ccfir,zhang2013cotscfi}; \tool{} applies
that motivation to CUDA device binaries whose executed ISA is SASS.

\subsection{Why CPU CFI Does Not Directly Transfer}

A conventional CPU-CFI workflow builds a CFG, derives target sets, protects
returns, inserts checks, and counts protected edges. CUDA/SASS weakens the
evidence behind each step. The executed code is post-lowering SASS. Target
evidence may be missing from a closed binary. Return state may be
register-backed, spilled, or optimized away. SIMT execution can make one static
transfer warp-uniform in one execution and lane-local in another. A CUDA binary
CFI system must classify recovered SASS forms first: calls and returns carry
continuation state, indirect transfers require target recovery, predication and
SIMT state determine check granularity, and termination defines fail-closed
behavior. Coverage gaps must be reported outside protected coverage.

\subsection{Return-State Exposure Is Not a CPU Stack}

The backward edge is especially difficult to analyze. Optimized CUDA code may
inline device functions, keep continuation state in registers, or remove
apparent source-level calls. A source-level helper call may disappear entirely
after inlining; another call may leave a visible \texttt{RET}, but with
continuation state kept in registers; only some lowering choices materialize a
writable local-memory continuation slot. A useful memory-backed backward-edge
exploit therefore requires: (1) continuation state, (2) writable local-memory
materialization, (3) attacker-controlled write reachability, (4) epilogue reload,
and (5) \texttt{RET} consumption. \texttt{RET} count is surface, not
exploitability. These requirements motivate a staged view of return exposure:
return-free code, register-backed returns, memory-backed static candidates, and
runtime-confirmed diversions should not be collapsed into a single count of
\texttt{RET} instructions.

This exposure is why \tool{} measures return-state exposure. A binary may have no
device-side returns, may have returns whose state is register-backed, may expose
a static memory-backed spill/reload candidate, or may contain a write primitive
that can reach and redirect that state. These cases imply different security
scope and different evaluation evidence.

\subsubsection*{Implications for CUDA binary CFI}

These observations define \tool{}'s scope. A CUDA binary defense must operate at
the native SASS boundary, often without source code. It must distinguish
no-surface code, fixed edges, recoverable indirects, unsupported sites, and
profile exclusions. It must authenticate dynamic continuation state for
protected backward edges and validate per-site target sets for recoverable
forward edges. Finally, it must treat SIMT granularity as part of the security
policy, because a transfer can be lane-local even when its static instruction is
shared by a warp. The rest of the paper implements and evaluates that
recovered-site policy.

\tool{} can reject forbidden or wrong-site targets, but same-site legal target
substitution remains residual surface \cite{carlini2015control}. GPU
memory-safety systems are complementary
\cite{gpushield2022,guardian2024,gpuarmor2025}.

\section{Threat Model and Security Property}
\label{sec:threat-model}

\tool{} assumes a trusted host/CUDA stack running non-malicious but vulnerable
CUDA device code. The attacker controls inputs and may obtain device-memory
corruption, including corruption of data later consumed as control flow.
\tool{} stops corrupted control-flow state at consumption time for recovered
protected SASS sites. Memory safety, code integrity, malicious-binary handling,
and full transfer coverage are separate problems.

\subsubsection*{Deployment and attacker model}

\tool{} targets GPU-backed applications and services that launch CUDA kernels,
framework extensions, or precompiled CUDA libraries on attacker-influenced
inputs. This setting matches recent CUDA exploitation work in which a trusted
host program launches vulnerable device code \cite{guo2024gpu,jacobs2025cuda}.

The protected CUDA binary is non-malicious: it may contain memory-corruption
bugs, but it is not written to evade \tool{}. \tool{} loads with the host
application, recovers SASS control-flow sites, generates a matching policy, and
places checks before protected kernels execute. Opaque or precompiled code is
protected only where its SASS evidence is recoverable.

The attacker controls inputs, triggers device-side memory corruption, and may
corrupt ordinary GPU memory reachable by the vulnerable kernel. This includes
global, local, or shared memory; function pointers; dispatch tables; jump-table
data; continuation slots; telemetry; and writable metadata. The attacker can
also cause denial of service by triggering fail-closed violations.

The attacker is not able to compromise the host process, CUDA driver/runtime, GPU firmware, selected backend, policy generator, protected binary after policy
generation, or backend-private state. Device read/write bugs are in scope; key
or backend-private disclosure is not. \tool{} authenticates writable metadata
under trusted backend-private state; it is not a GPU-resident secret-protection
mechanism.

\subsubsection*{Trusted boundary}

Ordinary CUDA memory may be corruptible and is treated as attack input. The
attacker may corrupt writable records, but writable records do not authorize
control flow. This includes writable shadow records, forward records,
telemetry, counters, and application buffers. Telemetry is diagnostic only and
never authorizes a transfer. State that can authorize a transfer is
authenticated, checked against trusted policy state, or kept in a
backend-private path. Device-readable application memory is attacker controlled.
Keys, helper-private arguments, and patch-cache private state are trusted
backend state. If a deployment exposes them to GPU-side reads, it needs driver,
runtime, or hardware isolation. Table~\ref{tab:app-trust-boundary} summarizes
this boundary.

\subsubsection*{Protected-site CFI property}
 For every SASS control-flow consumption site that
\tool{} classifies as protected, \tool{} checks the consumed return continuation
or indirect target before the transfer is released. If the value does not match
authenticated backward state or the recovered per-site forward target set,
enforcement fails closed. Unsupported, fallback, no-surface, fixed-edge, and
profile-excluded sites stay outside protected coverage.

This property is site-local. It defines what happens at a protected SASS
transfer. Every other source-level call, SASS transfer, or binary outcome stays
in the accounting. A deployment can reject binaries with unsupported indirect
sites, run them only in diagnostic mode, or accept the residual surface.
\tool{} makes that choice explicit instead of silently treating unknown SASS as
allowed control flow.

\subsubsection*{Goals}
\tool{} gives CUDA binary CFI a precise enforcement unit. For protected returns,
the consumed continuation must match authenticated expected state for that
thread/lane context. For protected indirect calls or branches, the dynamic
target must belong to the recovered per-site target set. Telemetry, counters,
and violation records never authorize transfers; return tokens and
forward-target records are authenticated or checked against trusted policy state
before use. Enforcement mode is the prevention mode. Diagnostic and detect-only
modes record violations but do not prevent release.

\subsubsection*{Non-goals}

\tool{} does not provide memory safety, code integrity, malicious-binary
handling, host/driver/firmware protection, side-channel defense, software-only
secret storage, complete SASS ISA coverage, vendor-agnostic GPU CFI, or
attack-prevalence measurement. Return-free, inlined, unsupported,
profile-excluded, and non-recoverable sites are reported outside protected
coverage.

\begin{figure*}[t!]
\centering
\small
\begin{tikzpicture}[
  font=\sffamily\scriptsize,
  >=Latex,
  lane/.style={draw=black!18, fill=black!1, rounded corners=4pt},
  box/.style={draw=black!55, fill=black!3, rounded corners=2pt, align=center,
    inner xsep=4pt, inner ysep=3pt, minimum height=0.62cm, text width=2.05cm},
  hostbox/.style={box, draw=black!45, fill=black!2},
  policybox/.style={box, draw=black!75, fill=black!8},
  gpubox/.style={box, draw=black!65, fill=black!5},
  failbox/.style={box, draw=black!85, fill=black!12, inner ysep=5pt,
    minimum height=0.82cm},
  auditbox/.style={draw=black!75, fill=black!6, rounded corners=2pt, align=center,
    inner xsep=4pt, inner ysep=3pt, minimum height=0.66cm},
  arr/.style={->, line width=0.55pt, draw=black!65, rounded corners=3pt,
    shorten <=1pt, shorten >=1pt},
  branch/.style={->, line width=0.55pt, draw=black!65, rounded corners=6pt,
    shorten <=0pt, shorten >=1pt},
  gatearr/.style={-{Latex[length=1.7mm,width=1.2mm]}, line width=0.55pt,
    draw=black!65, rounded corners=6pt, shorten <=1pt, shorten >=2pt},
  rail/.style={line width=0.55pt, draw=black!65, rounded corners=6pt},
  dataarr/.style={->, line width=0.45pt, draw=black!50, dashed,
    rounded corners=3pt, shorten <=1pt, shorten >=1pt},
  runtimearr/.style={-{Latex[length=1.5mm,width=1.05mm]}, line width=0.45pt,
    draw=black!50, dashed, rounded corners=3pt, shorten <=2pt, shorten >=2pt},
  lab/.style={font=\sffamily\bfseries\scriptsize, text=black!72, align=center}
]
\draw[lane] (-1.65,1.92) rectangle (2.05,-1.90);
\draw[lane] (3.05,1.92) rectangle (7.45,-1.90);
\draw[lane] (8.05,1.92) rectangle (15.75,-2.92);

\node[lab] at (0.20,1.56) {Host / load time};
\node[lab, text width=2.70cm] at (5.25,1.56) {Policy +\\check placement};
\node[lab] at (11.90,1.56) {GPU runtime};

\node[hostbox] (module) at (0.20,0.95) {\textbf{CUDA module}\\cubin/fatbin/JIT};
\node[hostbox] (recover) at (0.20,-0.10) {\textbf{SASS recovery}\\sites + targets};
\node[policybox] (policy) at (0.20,-1.15) {\textbf{Site outcomes}\\protected / gaps};

\node[policybox] (bundle) at (5.25,0.74) {\textbf{Bound policy}\\digest, targets, key use};
\node[hostbox, text width=3.10cm, minimum height=1.12cm] (rewrite) at (5.25,-1.18)
  {\textbf{Backend contract}\\WG-NVBit checks\\WG-ST timing patches\\WG-PC patch-cache};

\node[gpubox] (image) at (11.90,0.86) {\textbf{Checked SASS}\\instrumented or patched};
\node[gpubox] (back) at (9.75,-0.55) {\textbf{Backward check}\\return token};
\node[gpubox] (fwd) at (14.05,-0.55) {\textbf{Forward check}\\target set};
\node[failbox] (gate) at (11.90,-2.20) {\textbf{Gate / report}\\release, trap, record};

\node[auditbox, text width=3.00cm] (auditCoverage) at (0.20,-3.72)
  {\textbf{Coverage map}\\protected / fixed / gaps};
\node[auditbox, text width=2.45cm] (auditPolicy) at (5.25,-3.72)
  {\textbf{Policy record}\\digest + profile};
\node[auditbox, text width=3.25cm] (auditRuntime) at (11.90,-3.72)
  {\textbf{Runtime record}\\counters + first violation};

\coordinate (split) at (11.90,0.22);
\coordinate (merge) at (11.90,-1.18);

\draw[arr] (module) -- (recover);
\draw[arr] (recover) -- (policy);
\draw[arr] (policy.east) -- ++(0.45,0) |- (bundle.west);
\draw[arr] (bundle.south) -- (rewrite.north);
\draw[arr] (rewrite.east) -- ++(0.42,0) |- (image.west);
\draw[rail] (image.south) -- (split);
\draw[branch] (split) -| (back.north);
\draw[branch] (split) -| (fwd.north);
\draw[rail] (back.south) |- (merge);
\draw[rail] (fwd.south) |- (merge);
\draw[gatearr] (merge) -- (gate.north);

\draw[dataarr] (policy.south) -- (auditCoverage.north);
\draw[dataarr] (rewrite.south) -- (auditPolicy.north);
\draw[runtimearr] (gate.south) -- (auditRuntime.north);
\end{tikzpicture}
\caption{\tool{} architecture. Host-side analysis recovers SASS sites and binds
policy to the loaded image. Check placement is backend-specific: WG-NVBit places
reference helper checks, WG-ST places matched timing patches, and WG-PC loads
verified patch-cache entries for supported \texttt{sm\_89} surfaces. Runtime
checks enforce supported sites; unsupported and profile-excluded sites are
audited and excluded from protected CFI coverage.}
\label{fig:warpguard-pipeline}
\end{figure*}
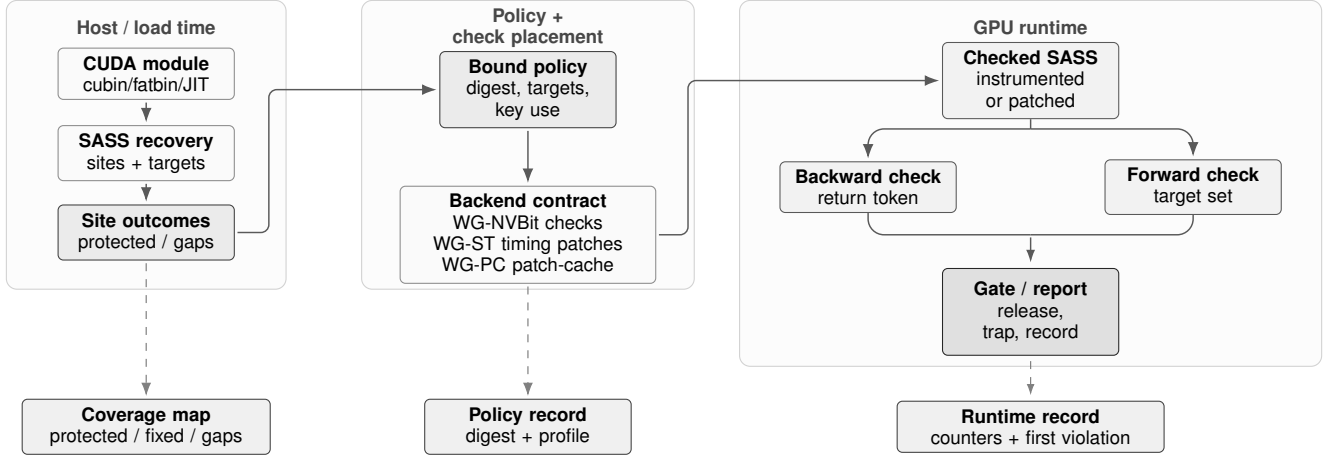

\section{System Design}
\label{sec:system-design}

\subsection{Overview}

\tool{}'s design follows four invariants. \textbf{SASS-boundary:} policy is
derived from executed SASS, not source or PTX intent. \textbf{Check-before-release:}
a protected transfer is not released before validation. \textbf{Authorization:}
telemetry and writable records never authorize a transfer without authenticated
or trusted policy state. \textbf{Accounting:} unsupported, fallback,
profile-excluded, fixed-edge, and no-surface outcomes stay outside protected
coverage.

Figure~\ref{fig:warpguard-pipeline} shows the solution overview. \tool{} is a
load-time CFI layer for CUDA device binaries: it recovers SASS, generates
policy, places checks or patch plans before protected execution, and emits an
audit record that separates protected sites from fixed-edge, excluded,
unsupported, and violated sites. The input is a cubin, fatbin, or JIT-produced
device image; the output is a checked execution path or bounded patch plan with
policy metadata bound to the recovered image.

\tool{} keeps the protected-site policy separate from check placement. WG-NVBit
is the broad security reference for corpus recovery and attack evaluation.
WG-ST is a matched static-trampoline timing lane for comparing callback-free
SASS placement with native and NVBit execution on the same \texttt{sm\_89}
timing cases. WG-PC is the reviewed patch-cache prevention lane for manifest-backed
\texttt{sm\_89} surfaces. These names denote scoped evidence lanes, not
different CFI policies: each backend must bind policy to recovered SASS, check
before release, preserve valid behavior, fail closed on invalid or unsupported
plans, and keep telemetry outside authorization.

\subsection{Architecture and Data Flow}

The architecture has four paths. The host path recovers SASS and synthesizes
policy. The placement path inserts NVBit helpers or emits static patch plans.
The runtime path checks supported transfers before release. The audit path
records coverage and violations. Authorization uses policy and runtime checks,
not audit counters; uncertain recovery becomes unsupported or fallback rather
than an implicit allow-any edge.

\subsection{Site Recovery and Policy Outcomes}

Policy synthesis uses SASS sites rather than source functions. A site is a SASS
instruction or sequence that consumes control-flow state: returns consume
continuations, indirect calls and branches consume targets, and direct calls may
create continuations even though their call target is fixed.

Policy generation assigns each site one outcome. Table~\ref{tab:site-outcomes}
defines the outcomes as part of the security design. Seeing a site is not enough to count it as protected.

\begin{table}[h]
\centering
\caption{\tool{} site outcomes}
\label{tab:site-outcomes}
\scriptsize
\setlength{\tabcolsep}{3pt}
\begin{tabular}{p{0.19\linewidth}p{0.38\linewidth}c p{0.21\linewidth}}
\toprule
Outcome & Runtime behavior & Protected? & Action \\
\midrule
protected & check inserted under the selected profile & yes & enforce \\
fixed-edge & fixed edge under the code-integrity assumption & no & account \\
unsupported & edge exists, but target or semantic evidence is insufficient & no & reject/diagnose \\
profile-excluded & protectable in principle, but disabled by the chosen mode & no & configure \\
no-surface & no dynamic CFI transfer is present & no & account \\
fallback & artifact, function, or site could not be analyzed safely & no & reject or fail closed \\
\bottomrule
\end{tabular}
\end{table}

The outcomes are deployment inputs. A strict deployment can reject unsupported
indirect sites; a diagnostic run can record them. Unsupported sites stay outside
protected coverage. Treating an unresolved indirect as ``allow any'' would
remove the security meaning of the forward policy.

\subsection{Authenticated Backward-Edge CFI}
Backward-edge CFI protects instrumented SASS returns at the point where the
continuation is consumed. On each protected call, \tool{} records the expected
return continuation, callsite identifier, depth, logical thread slot, and a keyed
token; profiles may also bind a launch epoch or monotonic counter. On return,
the helper reads the continuation that the SASS site is about to release and
verifies it against the authenticated record. A match releases the return; a
mismatch records a violation and fails closed.

The keyed token makes writable shadow metadata non-authoritative:

\begin{wgcode}
push(s, ret, slot, d):
  shadow[slot][d] = (ret, s, d, MAC_k(ret,s,slot,d,epoch))
check_ret(s, obs, slot, d):
  r = shadow[slot][d]
  require verify_MAC_k(r) && obs == r.ret
  release RET or trap
\end{wgcode}

Corrupting the return target, expected continuation, depth, or token does not
authorize a return unless the attacker can disclose trusted key material or
compromise the trusted host/runtime stack. Metadata faults such as underflow,
overflow, thread-slot overflow, and depth mismatch stop at the check instead of
aliasing another thread's state.

This protection applies only where a dynamic return exists and the selected
profile instruments it. Fully inlined code and return-free kernels have no
backward edge to check; visible returns can also be unsupported or
profile-excluded, and are reported as coverage gaps rather than protected sites.

\subsection{Selective Forward-Edge CFI}

Forward-edge CFI protects recoverable indirect calls and branches. The policy
is per-site. For each supported site, \tool{} stores an allowed target set. The
set may come from recovered binary metadata, address-taken device functions, or
dispatch-table/data-object analysis.

The supported-site set counts target
sets recovered by the policy generator; controlled evaluation policies, if
used, are reported separately and are not mixed into the recovered target-set
numerator. Sites with missing or ambiguous target evidence receive an
unsupported outcome instead of a coarse allow-any policy. Before the indirect
transfer is released, a helper checks that the observed target is in that
site's allowed set.

A typical protected forward edge is a device dispatch table or function
pointer. At load time, \tool{} binds the indirect SASS site to its recovered
target set. At runtime, the helper reads the dynamic target before release. A
target in the set is released; a forbidden target, corrupted table entry, or
target valid only at another site records a forward violation and fails closed.

\tool{} does not guarantee semantic dispatch integrity. If
two targets are both legal for the same site, \tool{} leaves that semantic
choice to the program or a stronger policy layer. That stronger property
requires semantic policy, control-data isolation, SFI, or compiler support.
\tool{}'s forward-edge policy is selective target-set CFI for recoverable sites.

\subsection{Metadata and Policy Binding}

\tool{} separates observability from authorization. Counters, telemetry buffers,
and first-violation records help audit coverage and failures, but they never
authorize a transfer; corrupting them can affect reports at most. Enforcement
decisions depend only on bound policy tables, authenticated return records,
forward target records, and runtime key material.

Protected configurations bind policy to the analyzed binary metadata. A binding
mismatch becomes an abort, unsupported outcome, or fallback rather than a wider
allowed target set. Measurement-only profiles may omit this binding and are not
prevention configurations. Valid keyed state authorizes checks under the trusted
host/runtime and instrumentation boundary; current GPUs do not provide a
hardware-protected secret for this software backend, so deployments that expose
backend-private state need stronger isolation.

\subsection{SIMT-Aware Validation}

SIMT execution fixes check granularity. A site can be warp-uniform or
lane-local; different lanes may carry different targets or return state.
\tool{} therefore uses conservative lane/thread-local checking by default. A
per-warp check is only an optimization after proving uniformity. The evaluation
includes a divergent-lane backward hijack to test that one corrupted lane cannot
hide inside a warp-level aggregate.

Predication and reconvergence are modeled separately from ordinary CFI edges.
Reconvergence instructions and barriers affect which lanes execute later
instructions. \tool{} instruments the sites that consume return or
indirect-target state and reports SIMT/control forms outside its safe model.

\subsection{Violation Handling}

\tool{} uses separate profiles for observation and prevention. Detect-only
profiles record violations and let the run continue. Enforcement profiles trap
before the invalid protected transfer is released. Fast or attribution profiles
that skip kernels, sites, counters, or a CFI direction change the protected set
and are reported under separate denominators; Table~\ref{tab:app-runtime-profiles}
lists the profile table used by the measurement scripts. After a failed control-flow
check, enforcement traps instead of trying in-kernel recovery, because the
execution state has already crossed a corruption boundary.

\section{Implementation}
\label{sec:implementation}

% This section describes the mechanisms needed to name SASS sites, place checks,
% represent runtime metadata, bind policy state, and report violations. Runtime
% costs are evaluated in Section~\ref{sec:optimization}.

\subsection{Check-Placement Backends}

The implementation realizes the same protected-site property through three
check-placement lanes. WG-NVBit uses NVBit as its SASS instrumentation substrate
\cite{villa2019nvbit}, loads into the host process, observes CUDA module
launches, and inserts helper checks into native SASS without source or compiler
changes. It is the broad reference backend for corpus recovery and attack
evaluation.

WG-ST is the static-trampoline timing lane. It preserves the final-SASS boundary
for a matched \texttt{sm\_89} timing surface, rewrites only those timing cases,
and exists to compare native, NVBit, and callback-free static execution on the
same kernels. WG-ST timing is not generalized to WG-PC prevention or to broad
corpus coverage.

WG-PC is the patch-cache SASS backend. Before module load, it selects reviewed
manifest-backed \texttt{sm\_89} patch plans, including checked, stateful
checked, callee-clone, and scoped dynamic-shadow surfaces. A manifest names the
architecture, module digest, matched SASS byte window, site identifiers, rewrite
form, scratch-register contract, predicate contract, expected policy digest,
and supported-surface label. WG-PC never generalizes a patch plan from one SASS
window to another unless this manifest match succeeds. Unmatched manifests,
scratch pressure, predicate mismatch, or unsupported geometry produce refusal
or fail-closed behavior rather than widened coverage. Table~\ref{tab:app-rewrite-forms}
summarizes the rewrite forms.

\subsection{Module and Site Discovery}

\tool{} recovers a CFI-oriented SASS IR from \texttt{nvdisasm} listings plus
optional \texttt{cuobjdump} ELF/CUDA metadata: functions, PCs, predicates,
control-transfer classes, return and indirect sites, and target evidence.
Ambiguous continuation or target evidence becomes unsupported, not allowed. At
launch, WG-NVBit maps the active CUDA module to reachable device functions;
callback-free lanes use file-backed cubins, patch plans, and verified
patch-cache metadata. The implementation walks each recovered function once,
unless reinstrumentation or patch regeneration is forced, and assigns stable
site identifiers to policy-relevant control transfers.

A site identifier is local to a recovered SASS instruction and its policy class:

\begin{wgcode}
site_id = H(module_digest || sm_arch || function_start ||
            sass_pc || site_class)
\end{wgcode}

\noindent The module digest binds the policy to the loaded image,
\texttt{sm\_arch} separates architecture-specific encodings,
\texttt{function\_start} and \texttt{sass\_pc} identify the instruction, and
\texttt{site\_class} separates return, indirect-call, indirect-branch, and
accounting-only sites. Returns receive backward-edge site identifiers. Indirect
calls and indirect branches receive forward-edge site identifiers. Direct
branches are recorded for coverage accounting but normally do not receive a
forward target check. The target  of direct calls is
fixed, but they may still create a return continuation, so they can receive
backward push instrumentation.

If a site exists but its target set, operands, or semantics cannot be recovered
with enough confidence, the implementation emits an unsupported record and gives
no protected-count credit. This rule is enforced in policy tables and runtime
audit output.

\subsection{Control-Transfer Instrumentation}
\tool{} checks just before any SASS transfer that consumes
control-flow state. WG-NVBit emits helper-call checks, while WG-ST and WG-PC
install SASS patches when the rewrite shape permits. Direct and indirect
\texttt{CALL}s execute \texttt{push(site, expected\_ret, guard)} before the
callee; \texttt{RET}s execute \texttt{check\_ret(site, observed\_ret, guard)}
before release; and indirect forward transfers materialize the target, validate
it against the site policy, and release the original transfer only when allowed
or in detect-only mode. Covered forward forms are indirect \texttt{CALL},
\texttt{BRX}, \texttt{BRXU}, \texttt{JMX}, and \texttt{JMPX}.

Checks preserve the original SASS predicate semantics. The backend receives or
reconstructs the predicate guard, because a predicated instruction may execute
for only a subset of active lanes. For \texttt{@P0 RET}, lanes with
\texttt{P0=false} do not execute the return and must not be checked as if they
had consumed a continuation. Active lanes check their own observed target or
continuation. WG-NVBit passes the guard predicate into its helper; WG-PC accepts
only rewrite shapes whose predicate and scratch-register behavior have been
validated for the supported surface.

The check also receives or materializes the values needed to identify the site
and observe the consumed control-flow value. Returns use the return-target
register state available at the return site. Indirect forward transfers use the
register-backed target operand or recovered branch-target representation. These
values are checked against policy metadata; they are not trusted by themselves.
Because SASS mnemonics and operands vary by architecture and compiler output,
the implementation normalizes site and operand classes before policy generation.
If the observed target or expected continuation cannot be materialized, the site
becomes unsupported or fallback. Table~\ref{tab:app-impl-support} lists the
runtime facts and per-form support boundary.

\subsection{Forward Target-Set Recovery}

The policy generator builds forward target sets before runtime instrumentation.
It first consumes target evidence already present in the CUDA binary utilities:
constant relocations, \texttt{nv.info} indirect-branch metadata, and
\texttt{nvdisasm} branch-target annotations when available. It then scans
recovered ELF/data objects that look like device dispatch tables or
address-taken function tables and resolves entries to recovered device
functions. A site is supported only when the observed indirect transfer can be
bound to one of these target sets. If the transfer is present but the target
evidence is missing, ambiguous, or unsupported by the current parser, the
generator emits an unsupported entry instead of a broad allow-any set.
Ambiguous evidence serves as a security decision: it becomes unsupported rather than allow-any.

Controlled evaluation policies are kept out of the recovered target-set
count. The provenance table reports binary-metadata, ELF/data-object, unknown,
evaluation-supplied negative-test, and unsupported entries separately. This keeps
the forward-CFI result tied to recovered evidence rather than to manual target
lists. Section~\ref{sec:evaluation} reports the full provenance audit in the
forward target-set table.

\subsection{Runtime Authorization State}

Backward-edge enforcement uses one logical shadow stack per CUDA thread slot.
The implementation linearizes the CUDA block and thread identifiers into a
\texttt{thread\_slot}, then indexes bounded per-thread depth arrays. Each
shadow entry stores \texttt{expected\_return}, \texttt{callsite\_id},
\texttt{depth}, and \texttt{token} for the current slot. The linearization uses
the launch geometry recorded for the protected kernel; if a launch exceeds the
configured capacity, the backward check fails closed instead of wrapping. The
scoped WG-PC dynamic-shadow manifests use the reviewed manifest capacity for
their supported geometries. Evaluated protected cases stay within configured
capacities; capacity failures are reported as boundary/fail-closed cases, not
compatibility passes.

The token is a keyed authenticator over the return state:

\begin{wgcode}
ret_token = MAC_k(expected_return, callsite_id,
                  depth, thread_slot, launch_epoch)
\end{wgcode}

\noindent The current prototype needs unforgeability under the trusted-key assumption;
writable shadow memory may remain ordinary writable device memory.

The push check computes the expected return continuation for the callsite and
writes the authenticated record at the current depth. The return check reads the
continuation that the \texttt{RET} will consume, reads the record at the current
depth, verifies the token, and compares the authenticated state with the
observed continuation. A valid return decrements the depth and releases the
transfer. A mismatch records a violation and, in enforcement mode, traps instead
of releasing the return normally.

Bounds failures are security failures. Thread-slot overflow, shadow-stack
overflow, underflow, and depth mismatch all fail closed instead of aliasing
another lane or thread.

Forward-edge enforcement uses per-site target records. A supported forward site
has a compact runtime record containing \texttt{site\_id}, \texttt{token},
\texttt{target\_count}, and \texttt{allowed\_targets}. The forward check receives
the site identifier and observed target, authenticates the record when metadata
integrity is enabled, and checks membership in the allowed target set. The
forward metadata token is a MAC over \texttt{site\_id},
\texttt{target\_count}, and \texttt{allowed\_targets}. If the target is absent,
the helper records a forward violation and traps in enforcement mode.

The target set is site-specific. A target may be a valid function address
elsewhere in the module and still be rejected because it is not valid for this
site. If two targets are both legal for the same site, the implementation
permits either target; that is the same-site semantic boundary.

\subsection{Keys and Policy Binding}

The implementation keeps authorization state in trusted host/runtime state. In
WG-NVBit, keys are passed to injected helpers as instrumentation arguments and
are not exposed as ordinary target-writable CUDA objects. In WG-PC, the trusted
host/runtime selects verified patch-cache entries and policy-bound metadata
before loading the protected module. The prototype derives process-local keys
from host entropy; a deterministic seed exists only for debugging and
reproducibility.

Backward tokens bind the expected return, callsite identifier, depth, logical
thread slot, and key material; some helper profiles also bind a launch epoch.
Forward metadata tokens bind the site identifier, target count, and allowed
target list. These tokens let the runtime store authorization records in
ordinary GPU memory without treating those records as self-authenticating.

Policy binding is checked before use. When an enforcement profile requires a
policy digest, a mismatch aborts before the policy is used. For static lanes, a
patch-cache entry is used only if its expected SASS bytes, policy digest,
architecture, and manifest entry match the module; otherwise WG-PC refuses the
patch plan and the site stays outside callback-free protected coverage. Ordinary
device symbols would be writable under the attacker model, so the security
result still depends on the threat-model assumption that a device bug cannot
read or corrupt backend-private arguments, patch-cache private state, or host
runtime state. The prototype assumes trusted backend-private key material; it
does not claim GPU-resident secret confidentiality against arbitrary device
disclosure.

\subsection{Violation Handling}

The runtime distinguishes observation from enforcement. In detect-only mode, a
failed check records counters and the first violation record, then lets the
kernel continue. In enforcement mode, the helper traps before the invalid
protected edge is released. The kernel fails closed at the CUDA boundary, and
the host can discard outputs, report the request as failed, or reset the CUDA
context.

Only authorization state controls edge release. Counters, telemetry buffers, and
first-violation records are diagnostic; corrupting them can affect logs but
cannot authorize a transfer. The runtime records the first violation because
that record is enough for a fail-closed decision.

\section{Evaluation}
\label{sec:evaluation}

We evaluate WG-NVBit, the broad reference backend, on four security points:
SASS recovery, protected-site coverage, enforcement on corrupted control state,
and residual legal or unsupported surface. Runtime cost and callback-free SASS
lanes use separate evidence: WG-ST for matched timing and WG-PC for reviewed
supported \texttt{sm\_89} prevention cases (Section~\ref{sec:optimization}).

\noindent\textbf{Security evidence.}
The evaluation checks six points: (1) whether \tool{} can recover SASS CFI
sites and assign explicit outcomes; (2) whether the protected denominator stays
separate from unsupported, no-surface, fallback, fixed-edge, and
profile-excluded cases; (3) whether backward checks stop corrupted
continuations before release; (4) whether forward checks stop forbidden or
wrong-site targets; (5) what legal or unsupported residual surface remains; and
(6) whether writable metadata, token, replay, and telemetry attacks fail closed
under the trusted-key boundary.

\noindent\textbf{Public-code evidence.}
Public-code cases then test whether the same protected-site abstraction appears
in real CUDA software.

\subsection{Setup and Evaluation Snapshot}

The main setup is an NVIDIA GeForce RTX 4070 Laptop GPU
(\texttt{sm\_89}, driver 581.83, CUDA 13.0) under Ubuntu-24.04. Remote
\texttt{sm\_75}, \texttt{sm\_86}, and \texttt{sm\_90} runs provide separate
attack-smoke, layout-probe, and atlas portability evidence
(Table~\ref{tab:app-sass-residuals}).

The corpus has six lanes: controlled/generated artifacts, source utilities,
standard/source roots, toolkit-library, external-binary, and Triton-cubin
inputs. They serve different roles, so recovery artifacts, policy entries,
dynamic checks, and attack cases are reported separately.

% The submitted artifact includes the clean \texttt{tool/} CLI, runbook,
% generated matrices, patch-cache manifests, and scripts for smoke checks and
% table inputs. GPU cases record CUDA, driver, architecture, NVBit, and manifest
% versions.

Each run emits static recovery, runtime counters, and first-violation records.
Attack cases compare native, detect-only, and enforcement behavior; benign cases
require correct output and no unexpected violation. The 77 recovered artifacts
and the 77 attack-matrix cases are different denominators.

Attack cases use a fixed three-run protocol. Native execution must reach the
attacker-selected behavior: marker landing, useful write, wrong target, or
metadata-forgery attempt. Detect-only execution must record the expected
violation and first-violation record. Enforcement must fail closed before
releasing the invalid protected edge.

\begin{tcolorbox}[colback=black!2,colframe=black!35,boxrule=0.25pt,arc=1pt,
  left=0.35em,right=0.25em,top=0.15em,bottom=0.15em,
  before skip=0.2em,after skip=0.2em]
\footnotesize
\textbf{Protected-evidence rule.} A case enters protected evidence only if
\tool{} recovers the SASS site, places an active check, observes corrupted
control state at that site, and fails closed in enforcement mode.
\end{tcolorbox}

Unsupported, fallback, no-surface, calibration, native-only, and
profile-excluded cases form explicit boundary cases.

Table~\ref{tab:eval-snapshot} summarizes the main denominators. The parser sees
51,621 final-SASS control-flow sites, 1,343 returns, and 160 indirect forward
sites. Protected counts require recovered policy evidence and active checks.

\textbf{Result.} \tool{} classifies 77 local artifacts, records 52.2M checked
edges, and separates recovered surface from protected enforcement evidence.

\begin{table*}[!t]
\centering
\caption{Evaluation snapshot and protected-site coverage. The table gives the
main local counts and the protected denominator used by the security
evaluation; corpus-lane and residual summaries are in
Tables~\ref{tab:app-corpus-lanes} and~\ref{tab:app-sass-residuals}.}
\label{tab:eval-snapshot}
\footnotesize
\setlength{\tabcolsep}{3pt}
\begin{tabular}{@{}p{0.24\textwidth}r p{0.54\textwidth}@{}}
\toprule
Metric & Value & Meaning \\
\midrule
Recovered artifacts & 77 & Artifacts whose SASS control-flow surface was classified. \\
Corpus lanes & 6 & Controlled/generated, source utilities, standard/source roots, toolkit-library, external-binary, and Triton-cubin lanes. \\
Classified SASS control-flow sites & 51,621 & Raw SASS recovery scale. \\
Returns & 1,343 & Backward-edge consumption surface. \\
Policy entries / protected artifacts & 138 / 31 & Sites or site groups that receive a CFI outcome. \\
Supported / unsupported forward entries & 154 / 15 & Recoverable target sets and explicit forward gaps. \\
Protected-active executions & 121 & Dynamic executions whose selected profile retains active protected checks. \\
Fallback/not-protected entries & 10 & Explicitly excluded from protected coverage. \\
Dynamic checks & 52.2M & 38.3M return checks and 13.9M forward checks. \\
Attack matrix & 77 cases & Native, detect-only, enforce, benign, boundary, and tamper cases. \\
Protected attack/benign cases & 52 & Cases counted as protected-site evidence. \\
\bottomrule
\end{tabular}
\end{table*}

Raw SASS sites, policy entries, forward target-set entries, dynamic executions,
and attack cases are different denominators. The 160 raw indirect forward sites are
instruction-level sites; the 154 supported and 15 unsupported forward entries are
policy/audit entries after target-set grouping and provenance splits. They are not
additive. Table~\ref{tab:app-evidence-reader} defines the terms used below.

\subsection{SASS Recovery and Protected-Site Coverage}

Recovery classifies each SASS control-flow site as protected, fixed-edge,
unsupported, profile-excluded, fallback, or no-surface. A \texttt{RET} is protected only if the selected profile
instruments it and the runtime can check the consumed continuation; an indirect
branch is protected only if its per-site target set is recovered. Direct
branches, return-free cubins, unsupported indirects, non-CFI
reconvergence/synchronization sites, and profile-excluded cases stay visible
outside protected evidence.

The protected set is nontrivial but smaller than recovered SASS surface. The
corpus has 1,343 returns and 160 indirect forward sites, but only sites with
binary evidence and an active protected profile enter the numerator. Recovery
validation passes 8/8 binary-utility checks and 6/6 source-known
function/return/call matches.

\textbf{Result.} \tool{} recovers enough SASS structure to derive enforceable
policy while refusing to count unknown sites as protected. The set contains 31
artifacts with protected sites and 121 protected-active executions. It
excludes 15 unsupported-site entries, 10 fallback/not-protected entries, return-free
code, fixed-edge transfers, and non-CFI reconvergence/synchronization surface.

\subsection{Forward Target Recovery and Residual Surface}

Forward evaluation checks two properties: forbidden targets must fail closed,
and the remaining legal target sets must be explicit. We first recover per-site
target sets from binary metadata and ELF/data-object evidence. We then overwrite
function pointers or dispatch-table entries to test forbidden targets and
targets valid only at another site. Finally, we count singleton target sets and
same-site alternatives that remain legal under target-set CFI.

Forward validation is selective. Supported target sets come from recovered
binary evidence; evaluation-supplied target sets are used only for controlled
negative tests. Table~\ref{tab:forward-merged} merges support, provenance,
precision, and residual same-site surface. Support and provenance entries use
different granularities.

\begin{table*}[!t]
\centering
\caption{Forward target-set recovery and precision. The first entries report the
site-level forward support boundary. The provenance entries are detailed audit entries
and use a different granularity from the supported-site count. Same-site legal alternatives
are residual semantic surface: \tool{} blocks forbidden and wrong-site targets,
but allows any target recovered for the current site.}
\label{tab:forward-merged}
\footnotesize
\setlength{\tabcolsep}{3pt}
\begin{tabular}{@{}p{0.36\textwidth}r p{0.40\textwidth}@{}}
\toprule
Metric & Value & Interpretation \\
\midrule
\multicolumn{3}{@{}l}{\textit{Site-level support boundary}}\\
Supported forward target-set entries (site-level) & 154 & Recoverable per-site target sets. \\
Unsupported forward audit entries & 15 & Explicit target-evidence gaps. \\
\addlinespace[2pt]
\multicolumn{3}{@{}l}{\textit{Detailed provenance audit}}\\
Detailed provenance audit entries & 163 & Lower-level audit entries used to explain evidence sources. \\
Automatic binary metadata entries & 114 & Detailed entries counted as recovered evidence. \\
Automatic ELF/data-object entries & 48 & Detailed entries counted as recovered evidence. \\
Evaluation-supplied entries & 0 & Not used in supported-site evidence. \\
Unknown or unattributed entries & 1 & Detailed entry audited separately. \\
\addlinespace[2pt]
\multicolumn{3}{@{}l}{\textit{Target-set precision}}\\
Median / p90 / max target-set size & 1 / 3 / 10 & Most target sets are small; some remain broad. \\
Median authority reduction vs. module functions & 71.4\% & Relative to allowing any recovered function in the module. \\
Single-target entries & 118 & No same-site residual alternative. \\
Multi-target same-site entries & 45 & Legal same-site target bending remains possible. \\
Small residual entries / broad residual entries & 39 / 6 & Residual target sets with 2--3 or more than 3 targets. \\
Residual same-site alternatives & 94 & Legal alternatives left by target-set CFI. \\
\bottomrule
\end{tabular}
\end{table*}

Wrong-site and forbidden targets are rejected.
Same-site valid-target substitution remains a semantic dispatch-integrity
problem. Table~\ref{tab:app-sass-residuals} summarizes the symbol-name residual
screen as supporting triage.

\textbf{Result.} \tool{} recovers 154 supported per-site target-set entries from
binary evidence. Relative to allowing any recovered function in the module, the
median supported site reduces target authority by 71.4\%. No
evaluation-supplied target sets enter that count. Forbidden and wrong-site
targets fail closed. The limit is 15 unsupported forward entries and 45 supported
entries with more than one legal target for the same site. Those 94 same-site
alternatives remain legal.

\subsection{Return Exposure and Backward Evidence}

The backward result separates return surface from return exploitability. Since
CUDA returns lack a conventional stack, we report observed return exposure
instead of treating every \texttt{RET} as equally exploitable. The audit tracks:
return-free code; register-backed returns; frames that do not yield a useful
writable continuation; static spill/reload candidates through writable memory;
and confirmed diversion. Confirmed diversion also requires dynamic evidence
that attacker-controlled state reaches the continuation consumed by a
\texttt{RET}.

The backward surface is large: the corpus has 1,343 returns and 9,727
return-terminated gadgets. The deployed-binary search reaches static
spill/reload candidates and shows why RET/gadget counts are not exploit counts:
confirmed natural diversion in third-party external code remains unobserved in
the evaluated corpus. The enforcement cases therefore test the consumption point
with controlled, source-assisted, seeded app/benchmark-rooted, and
prior-work-shaped cases. When a corrupted continuation reaches a protected
return, the authenticated check stops it.

This exposure progression is separate from attack realism. The L0--L4 ladder
classifies how each exercised evidence case is constructed. L0 cases are benign
checked paths, L1 cases corrupt one targeted control object, L2 cases compose
primitives such as useful writes, mini-chains, or divergent-lane returns, L3
cases use source-assisted ABI stress or seeded realistic kernels, and L4 cases
come from prior-work-shaped or public-code evidence. The backward L4 cases adapt
the published Guo et al. return-corruption shape to the local architecture
\cite{guo2024gpu}. The evaluation includes a local Guo-style primitive and an
architecture-adapted harness built from the pinned public PoC source,
preserving the vulnerable \texttt{sum1}/\texttt{sum2} device-function shape
under a stable marker landing. It reaches L4 while keeping return-exposure
evidence and attack-realism evidence separate.

\textbf{Result.} \tool{} reaches controlled, composed, source-assisted, seeded
realistic, and prior-work-shaped backward evidence.

\subsection{Attack Blocking}

The attack packet is a controlled 77-case matrix covering backward-edge
corruption, forward-edge target corruption, SIMT divergence, realistic
seeded kernels, prior-work shape, and metadata tamper. It measures whether
\tool{} stops corrupted control state once it reaches a protected SASS
consumption site.

Protected cases follow a fixed protocol: native execution reaches the
attacker-selected behavior, detect-only records the violation, and enforcement
fails closed before release. Boundary cases stay outside protected evidence:
same-site legal forward targets are allowed by design, while unsupported,
no-surface, and calibration cases remain explicit exclusions. Across protected
backward, forward, SIMT stress, seeded realistic, prior-work-shaped, and
metadata-tamper cases, \tool{} fails closed under the threat model.

\textbf{Result.} In protected cases, native execution reaches the attacker
behavior, detect-only records the expected violation, and \tool{} enforcement
fails closed before releasing the invalid edge.

\begin{table*}[!t]
\centering
\caption{Attack outcomes. The full 77-case regression matrix is summarized by
evidence class; Table~\ref{tab:app-security-evidence} gives grouped supporting
evidence and generated artifacts contain the per-case outputs.}
\label{tab:attack-outcomes}
\scriptsize
\setlength{\tabcolsep}{2.5pt}
\begin{tabular}{@{}p{0.21\textwidth}p{0.20\textwidth}p{0.16\textwidth}p{0.13\textwidth}p{0.20\textwidth}@{}}
\toprule
Class & Native behavior & Detect-only & Enforce & Scope \\
\midrule
Backward primitives & Return hijack, useful-write gadget, and two-return
mini-chain reach attacker behavior. & \texttt{ret\_viol}. & Fail closed. &
Controlled and composed return-consumption cases. \\
SIMT and ABI stress & Divergent-lane return and ABI-stress cases divert or land
markers natively. & \texttt{ret\_viol}. & Fail closed. & Lane-local and
compiler/ABI exposure cases. \\
Seeded realistic cases & App, benchmark, CUDA-sample-style, and ML-shaped cases
reach marker/value or wrong target. & \texttt{ret\_viol} or
\texttt{fwd\_viol}. & Fail closed. & Seeded bugs inside realistic kernel
structure. \\
Prior-work-shaped backward cases & Guo-style and upstream-derived adapted Guo
cases reach marker landing. & \texttt{ret\_viol}. & Fail closed. & Published
return-corruption shape adapted to local SASS. \\
Forward substitutions & Function pointer, dispatch table, and wrong-site valid
target reach forbidden behavior. & \texttt{fwd\_viol}. & Fail closed. &
Selective per-site target-set CFI. \\
Metadata and policy tamper & Token, policy, replay, and telemetry tamper
attempt corruption. & Expected violation for authorization tamper. & Fail
closed when authorization state is invalid. & Key-boundary red team. \\
\bottomrule
\end{tabular}
\end{table*}

%\subsubsection*{External Public-Code Evidence}

\subsection{External Public-Code Evidence}

Controlled cases show enforcement whereas public-code cases test whether \tool{}'s
protected-site abstraction appears in real CUDA software. We searched
applications, libraries, generated-code systems, and issue replays for state
that ordinary memory bugs can later turn into SASS control flow. Each case uses
linked \texttt{sm\_89} SASS recovery, protected/unsupported classification, and
a validation run.

The cases fall into six patterns: device dispatch tables (PPL-CUDA-SMC),
uploaded operation pointers (PNNL SV-Sim/DM-Sim), generated callable tables
(Dr.Jit), callback consumers (cuFFT users), runtime-initialized tables
(CUDA-Q, GooFit, Kokkos), and real return consumption
(GooFit issue~\#242).

PPL-CUDA-SMC consumes a device \texttt{pplFunc\_t} dispatch table in
\texttt{execFuncs}: corrupting it to an out-of-set marker prints
\texttt{marker=0x51c0} natively, records \texttt{fwd\_viol=1} in detect-only,
and exits before the marker under enforcement. PNNL SV-Sim and DM-Sim consume
uploaded \texttt{Gate::op} pointers; corrupting one outside the recovered
38-target family reaches markers natively, records forward violations, and
fails closed.

Other cases cover runtime-initialized tables, cuFFT callback consumers, generated
callable tables, and the GooFit issue~\#242 replay, which reaches protected
return consumption in real kMatrix/Thrust kernels. The rule is the same as in
the controlled matrix: recover or seal the target set, check before releasing
the SASS transfer, and fail closed for out-of-policy targets or continuations.
Public CUDA systems already contain device-resident control-flow objects whose
SASS consumption can be recovered and checked.
Table~\ref{tab:app-public-code-evidence} gives the case-level counts,
outcomes and upstream GitHub citations.

\subsection{Key-Boundary Validation}

The metadata-tamper cases test a common CFI failure mode: treating writable
records as authoritative. We corrupt return metadata, replay or tamper backward
tokens, corrupt forward and indirect-policy records, and tamper telemetry.
Matching detect/enforce cases record the expected violation and fail closed,
while telemetry corruption only changes diagnostics.

The packet also includes a writable-shadow-stack baseline. If both the real
continuation and an unauthenticated shadow copy are attacker-controlled, a
simple equality check can be forged. \tool{} instead verifies keyed state under
the trusted instrumentation boundary, validating tamper detection under the
declared key boundary.

Writable records are non-authoritative unless the keyed state verifies. The
result depends on the trusted backend boundary; disclosure or forgery of
backend-private key material, helper-private arguments, patch-cache private
state, or host runtime state requires stronger isolation.

\textbf{Result.} \tool{} rejects return-metadata corruption, token tamper,
token replay, forward-policy corruption, and telemetry tamper under the declared
key boundary. Key disclosure is a stronger-isolation problem.

\subsection{Compatibility and Portability Boundaries}

We test whether protected benign transfers still execute correctly. The packet
records 13/13 controlled benign protected cases and 7/7 CFI-active
standard-suite protected cases passing. Cases without counters, cases that fail
outside \tool{}, build-only cases, and benign-violation cases stay outside the
protected set.

Remote packs ran on an RTX 2080 Ti (\texttt{sm\_75}), an NVIDIA A10
(\texttt{sm\_86}), and an H100 80GB HBM3 (\texttt{sm\_90}). They validate
build, SASS recovery, representative backward/forward enforcement,
metadata-tamper detection, return-layout probes, and minimal corpus parsing.
These are representative portability checks, not full multi-architecture
reruns.

\textbf{Result.} \tool{} preserves the checked benign cases: 13/13 controlled
benign protected cases and 7/7 CFI-active standard-suite protected cases pass.
Remote architecture packs pass representative portability checks: 24/24
attack-smoke cases, 208/208 return-layout probes, and 36/36 minimal atlas cases.

\section{Backend Cost and Callback-Free Feasibility}
\label{sec:optimization}

\tool{} reports backend cost by evidence lane. WG-NVBit measures the broad
reference backend. WG-ST and WG-PC measure callback-free SASS paths only on
their supported surfaces. We evaluate four backend questions: the full WG-NVBit
reference cost, direct native/WG-NVBit/WG-ST/WG-PC timing on identical
\texttt{sm\_89} timing cases, WG-PC supported-manifest prevention, and WG-PC real-cubin
accepted-site breadth. Only the matched timing cases are direct runtime
comparisons across backends.

All headline timing measurements use the same RTX 4070 Laptop GPU setup as the
main evaluation. We report percent overhead for WG-NVBit end-to-end profiles
and runtime ratios for the matched callback-free timing packet. The runner uses
warmups, repeats, separate host wall-time records, and native-runtime buckets so
sub-millisecond launches do not dominate the reported ratios.

\begin{table}[!t]
\centering
\caption{Backend evidence firewall. Each evidence class supports only the claim shown in
the middle column.}
\label{tab:backend-evidence-firewall}
\footnotesize
\setlength{\tabcolsep}{2.5pt}
\begin{tabular}{@{}p{0.25\columnwidth}p{0.36\columnwidth}p{0.27\columnwidth}@{}}
\toprule
Evidence & Allowed claim & Not claimed \\
\midrule
WG-NVBit corpus & Broad reference enforcement. & Deployment overhead. \\
Attribution cases & CFI logic is not dominant after helper plumbing. &
Full-system cost. \\
Matched timing & Callback-free lanes avoid the NVBit callback floor. &
Speedup or production overhead. \\
WG-PC manifests & Reviewed \texttt{sm\_89} prevention. &
Arbitrary-binary replacement. \\
Real-cubin sweep & Supported sites exist in real cubins. &
General coverage. \\
\bottomrule
\end{tabular}
\end{table}

\subsubsection*{WG-NVBit: broad reference cost} WG-NVBit is the security reference
backend used for the broad corpus and attack matrix. As expected for a binary
instrumentation framework, it is very expensive:
protected workloads at or above 50 ms show 799.7--927.6\% overhead. Controlled
attribution separates that cost from policy logic. After helper plumbing,
metadata, token computation, and fail-closed checks each add below 1.3\%. The
result identifies the main cost source: dynamic instrumentation and helper
dispatch dominate the local reference backend, not the CFI predicate itself.

\subsubsection*{Matched timing: direct backend comparison} The direct
native/WG-NVBit/WG-ST/WG-PC comparison uses the same two \texttt{sm\_89} timing cases,
kernels, and attack mode. On that matched surface, WG-NVBit carries a
99.27\(\times\) native callback floor. WG-ST and WG-PC execute the same cases
without that callback floor; the detailed ratios are reported in
Table~\ref{tab:app-backend-evidence}. WG-PC falls below native on these short
static fixtures, and we treat that only as evidence that the NVBit callback
floor is absent, not as a speedup. In summary, callback-free SASS placement
removes the NVBit timing floor on the shared surface.

\subsubsection*{WG-PC: reviewed callback-free prevention} WG-PC is the reviewed patch-cache
prevention lane. It intercepts CUDA module loads,
selects verified patch-cache entries, emits checked \texttt{sm\_89} patches for
supported plans, and fails closed for unsupported modules or unsupported sites.
Its runtime regression packet uses repeat 30 and warmup 5, passes fail-closed
negative cases, and passes 13/13 dynamic-shadow validation cases. This shows
callback-free prevention on reviewed manifests. These cases are not pooled with
WG-NVBit overhead or WG-ST timing. Overhead for broader cases
remains a larger static-rewriting question.

\subsubsection*{WG-PC: real-cubin breadth} The real-cubin sweep tests whether the
patch-cache path finds protected sites outside hand-built fixtures. It accepts
374 protected sites across 17 of 59 selected CUDA sample, CUTLASS, Parboil, and
PolyBench cubins, and refuses cache entries whose manifest, SASS-window match,
or supported rewrite form fails validation instead of widening coverage. The
result is accepted-site breadth for reviewed \texttt{sm\_89}
cubins, bounded by architecture, manifest review, recovered target sets, and
semantic-preserving rewrite shapes. A separate benign-output packet exercises
three positive real cubins from that sweep: PolyBench-GPU ADI, CUDA Samples
\texttt{dxtc}, and CUDA Samples Function\-Pointers. In all three, the existing
application runner executes the original and WG-PC-patched cubins with the same
checksum, adding execution evidence for 31 accepted protected sites. The same
packet corrupts one \texttt{FunctionPointers} table entry and observes WG-PC
fail closed on the patched cubin. The native corrupted case faults rather than
reaching a useful attacker marker, so this is a selected invalid-target
fail-closed case, not a full attack case, and it does not generalize WG-PC beyond
reviewed patch-cache entries.

\section{Discussion}

\subsubsection*{Enforcement Semantics}

\tool{} protects recovered SASS consumption sites. A protected return checks the
observed continuation against authenticated dynamic state before release; a
protected indirect call or branch checks the observed target against the
recovered per-site target set. This is the unit of the security claim.
Unsupported/profile-excluded sites are coverage evidence, not protected edges.
Fallback, fixed-edge, and no-surface cases also stay outside the protected
denominator.

This distinction matters because writable metadata alone is not authority. A
writable shadow copy can be forged if the attacker corrupts both the real
continuation and the copy. \tool{} instead requires a keyed return record or
trusted policy state before releasing the transfer. Telemetry, counters, and
violation logs can explain a run, but never authorize control flow.

The public-code cases show that these shapes are not limited to fixtures. CUDA-Q,
GooFit, Kokkos, cuFFT callback users, Dr.Jit, PPL-CUDA-SMC, and PNNL
SV-Sim/DM-Sim contain device-resident control-flow objects consumed by SASS.
GooFit issue~\#242 also reaches protected return consumption in a public issue
replay.

\subsubsection*{Residual Attack Surface}

\tool{} checks consumption after a corrupting write. It does not make the write
safe, and it does not make failed-kernel outputs trustworthy. A deployment that
uses enforcement failures as a security signal should discard those outputs and
reset or quarantine the CUDA context. Bounds checking, temporal safety,
store-side SFI, and control-data isolation remain complementary.

Forward CFI blocks forbidden and wrong-site targets, but it allows any target
recovered for the current site. The evaluation reports 118 singleton entries, 45
multi-target entries, and 94 legal same-site alternatives. This is the standard
residual of set-based CFI~\cite{abadi2005cfi,carlini2015control}: target-set
membership does not prove semantic intent. Reducing that residual requires a
semantic policy, for example type/arity recovery, dispatch-table identity,
selector checks, table sealing, object provenance, isolation, or compiler
support.

Backward CFI is checked at return consumption. The external-binary search finds
static memory-backed candidates but no natural third-party case with confirmed
attacker reachability and native diversion. This is a measurement boundary, not
a claim that such cases cannot exist. CUDA/SASS backward exploitation needs
materialized continuation state, a reachable write, a live reload into the
return operand, and native diversion. \texttt{RET} and gadget counts are surface
metrics, not exploitability metrics.

Metadata tamper and replay cases exercise this boundary: writable
records may be corrupted, but release still depends on authenticated or trusted
backend-private state. If helper-private, patch-cache-private, or host/runtime
key material is exposed to GPU-side disclosure, deployments need driver,
runtime, or hardware isolation.

\subsubsection*{Deployment Choices}

%The accounting gives deployments several choices. 
Strict mode rejects
unsupported indirects, fallback sites, or profile exclusions. Diagnostic mode
records unsupported surface and continues. WG-NVBit is the 
backend when its cost is acceptable. WG-PC applies only to manifest-supported
\texttt{sm\_89} surfaces; unmatched modules must be rejected, diagnosed, or
handled by another backend. WG-ST is only the matched timing lane.

NVBit cost is dominated by dynamic instrumentation and helper plumbing.
Low-overhead deployment needs semantic-preserving SASS patching with the same
check-before-release contract. WG-PC shows this path for reviewed surfaces.
Arbitrary-binary replacement still needs manifest generation,
liveness/predicate validation, scratch-register selection, branch-window
checks, post-patch disassembly, and negative tests.

\subsubsection*{Portability and Hardening}

The main evaluation targets RTX 4070/\texttt{sm\_89}/CUDA 13. Remote
\texttt{sm\_75}, \texttt{sm\_86}, and \texttt{sm\_90} packs cover build,
recovery, selected enforcement, layout probes, and minimal atlas parsing. They
are sanity checks for architectural drift, not full multi-architecture reruns or
architecture-generic WG-PC support.

Static patch plans bind SASS bytes, architecture, policy digest, register use,
predicate semantics, and patch-window shape. New GPUs or \texttt{ptxas}
versions can change encodings, register allocation, call/return lowering,
branch windows, or scratch pressure; unmatched manifests fail closed. Broader
deployment needs multi-architecture rewriting, stronger patch validation,
hardware- or driver-backed key protection, and optional semantic dispatch
policy.

\section{Related Work}

\subsubsection*{CPU CFI and code reuse}
ROP, return-less ROP, JOP, and object-oriented code reuse showed that memory
corruption remains useful after code injection is blocked
\cite{shacham2007rop,roemer2012rop,checkoway2010noret,bletsch2011jop,
schuster2015coop}. CFI restricts such transfers to policy-approved edges
\cite{abadi2005cfi,burow2017cfisurvey}. Compiler and host-platform mechanisms
such as Clang CFI, ShadowCallStack, Microsoft CFG, Intel CET, ARM pointer
authentication, and RISC-V CFI protect CPU binaries or ISA-level host transfers
\cite{tice2014forwardcfi,clang_cfi,clang_shadowcallstack,microsoft_cfg,
intel_cet,arm_pac,riscv_cfi}. These mechanisms do not derive per-site policies
for user CUDA SASS instructions executing on NVIDIA SMs.

\subsubsection*{Binary CFI}
The closest CPU lineage is binary-level CFI. CCFIR and CFI for COTS binaries
showed that source-free CFI can use disassembly, relocation or layout evidence,
rewriting, and runtime mediation \cite{zhang2013ccfir,zhang2013cotscfi}.
PathArmor, TypeArmor, $\tau$CFI, and UCT studied context sensitivity,
argument/type recovery, and target-set precision
\cite{vanderveen2015patharmor,vanderveen2016typearmor,muntean2018taucfi,
hu2018uct}. Work on CFI attacks and metrics shows why coarse policies, legal
target bending, compatibility, and residual block utility must be measured
\cite{davi2014stitching,carlini2015control,evans2015controljujutsu,
conti2015losingcontrol,xu2019confirm,frassetto2022cfinsight}. \tool{} inherits
those evaluation concerns, but moves the policy and enforcement boundary to
executed CUDA SASS under SIMT execution.

\subsubsection*{GPU exploitation and memory safety}
CUDA overflow studies showed that GPU memory bugs can corrupt sensitive device
state \cite{di2015cudabuffer,di2016overflow}. Guo et al. demonstrated modern GPU
memory exploitation with code injection and code reuse, and Roels, Jacobs, and
Volckaert studied input-triggered device-side CUDA/SASS ROP
\cite{guo2024gpu,jacobs2025cuda}. Other GPU work shows isolation, ASLR, and
host-facing attack-surface risks \cite{dipietro2016cudaleaks,zhu2026gpuaslr,
roh2026ghost}. GPUShield, cuCatch, CuSafe, Guardian, GPUArmor, Compute
Sanitizer, and MIG address memory errors, instrumentation, or isolation
\cite{gpushield2022,cucatch2023,lu2026cusafe,guardian2024,gpuarmor2025,
nvidia_compute_sanitizer,nvidia_mig}. These systems reduce or contain the
corrupting memory event. \tool{} checks the later control-flow consumption when
corrupted return state or indirect targets are reached.

\subsubsection*{GPU TEEs and binary instrumentation}
Graviton, HIX, Telekine, Honeycomb, SAGE, and NVIDIA confidential-computing work
protect GPU execution, attestation, or confidential-computing boundaries
\cite{volos2018graviton,jang2019hix,hunt2020telekine,mai2023honeycomb,
ivanov2023sage,nvidia_confidential_computing}. GPU Ocelot, SASSI, NVBit,
NVBitFI, NVIDIA binary utilities, and NVLift provide PTX/SASS analysis,
instrumentation, fault injection, or lifting substrates
\cite{diamos2010ocelot,stephenson2015sassi,villa2019nvbit,tsai2021nvbitfi,
nvidia_ptx,nvidia_nvcc,nvidia_binary_utils,wan2026nvlift}. They are enabling
substrates or adjacent protections. \tool{} adds the SASS-level CFI policy:
classify sites, authenticate backward-edge state, validate recoverable forward
targets, fail closed, and report unsupported sites separately.

\section{Conclusion}

\tool{} applies protected-site CFI at the CUDA SASS boundary. Deployed
CUDA binaries do not admit a single whole-program CFI policy: some sites have
recoverable evidence, some are fixed, some lack dynamic surface, and some remain
unsupported. \tool{} enforces the boundary where evidence is sound:
authenticated backward-edge checks for instrumented returns, selective per-site
validation for recoverable forward edges, and fail-closed handling for invalid
protected transfers. On 77 CUDA artifacts, \tool{} classifies 51,621 SASS
control-flow sites, records 52.2M dynamic checks, and blocks representative
backward- and forward-edge attacks. Public-code cases confirm the same pattern in
real CUDA software, including dispatch tables, callbacks, generated callable
tables, and runtime-initialized tables. The result is an auditable enforcement
point for CUDA SASS sites where CFI can be derived and checked; to our
knowledge, it is the first evaluated CUDA device-binary CFI system at this
boundary.

\bibliographystyle{IEEEtran}
\bibliography{refs}

@inproceedings{guo2024gpu,
  author       = {Yanan Guo and
                  Zhenkai Zhang and
                  Jun Yang},
  title        = {{GPU} Memory Exploitation for Fun and Profit},
  booktitle    = {Proceedings of the 33rd {USENIX} Security Symposium ({USENIX} Security)},
  pages        = {4033--4050},
  publisher    = {{USENIX} Association},
  year         = {2024},
}

@inproceedings{jacobs2025cuda,
  author       = {Jonas Roels and
                  Adriaan Jacobs and
                  Stijn Volckaert},
  title        = {CUDA, Woulda, Shoulda: Returning Exploits in a SASS-y World},
  booktitle    = {Proceedings of the 18th European Workshop on Systems Security ({EuroSec})},
  pages        = {40--48},
  publisher    = {{ACM}},
  year         = {2025},
  doi          = {10.1145/3722041.3723099},
}

@inproceedings{carlini2015control,
  author       = {Nicholas Carlini and
                  Antonio Barresi and
                  Mathias Payer and
                  David A. Wagner and
                  Thomas R. Gross},
  title        = {Control-Flow Bending: On the Effectiveness of Control-Flow Integrity},
  booktitle    = {Proceedings of the 24th {USENIX} Security Symposium ({USENIX} Security)},
  pages        = {161--176},
  publisher    = {{USENIX} Association},
  year         = {2015},
}

@inproceedings{abadi2005cfi,
  author       = {Mart{\'{\i}}n Abadi and
                  Mihai Budiu and
                  {\'{U}}lfar Erlingsson and
                  Jay Ligatti},
  title        = {Control-flow integrity},
  booktitle    = {Proceedings of the 12th {ACM} Conference on Computer and Communications Security ({CCS})},
  pages        = {340--353},
  publisher    = {{ACM}},
  year         = {2005},
  doi          = {10.1145/1102120.1102165},
}

@inproceedings{shacham2007rop,
  author       = {Hovav Shacham},
  title        = {The geometry of innocent flesh on the bone: return-into-libc without
                  function calls (on the x86)},
  booktitle    = {Proceedings of the 14th {ACM} Conference on Computer and Communications Security ({CCS})},
  pages        = {552--561},
  publisher    = {{ACM}},
  year         = {2007},
  doi          = {10.1145/1315245.1315313},
}

@inproceedings{checkoway2010noret,
  author       = {Stephen Checkoway and
                  Lucas Davi and
                  Alexandra Dmitrienko and
                  Ahmad{-}Reza Sadeghi and
                  Hovav Shacham and
                  Marcel Winandy},
  title        = {Return-oriented programming without returns},
  booktitle    = {Proceedings of the 17th {ACM} Conference on Computer and Communications Security ({CCS})},
  pages        = {559--572},
  publisher    = {{ACM}},
  year         = {2010},
  doi          = {10.1145/1866307.1866370},
}

@inproceedings{bletsch2011jop,
  author       = {Tyler K. Bletsch and
                  Xuxian Jiang and
                  Vincent W. Freeh and
                  Zhenkai Liang},
  title        = {Jump-oriented programming: a new class of code-reuse attack},
  booktitle    = {Proceedings of the 6th {ACM} Symposium on Information, Computer and Communications Security ({ASIACCS})},
  pages        = {30--40},
  publisher    = {{ACM}},
  year         = {2011},
  doi          = {10.1145/1966913.1966919},
}

@inproceedings{tice2014forwardcfi,
  author       = {Caroline Tice and
                  Tom Roeder and
                  Peter Collingbourne and
                  Stephen Checkoway and
                  {\'{U}}lfar Erlingsson and
                  Luis Lozano and
                  Geoff Pike},
  title        = {Enforcing Forward-Edge Control-Flow Integrity in {GCC} {\&} {LLVM}},
  booktitle    = {Proceedings of the 23rd {USENIX} Security Symposium ({USENIX} Security)},
  pages        = {941--955},
  publisher    = {{USENIX} Association},
  year         = {2014},
}

@inproceedings{zhang2013ccfir,
  author       = {Chao Zhang and
                  Tao Wei and
                  Zhaofeng Chen and
                  Lei Duan and
                  Laszlo Szekeres and
                  Stephen McCamant and
                  Dawn Song and
                  Wei Zou},
  title        = {Practical Control Flow Integrity and Randomization for Binary Executables},
  booktitle    = {Proceedings of the 34th {IEEE} Symposium on Security and Privacy ({S\&P})},
  pages        = {559--573},
  publisher    = {{IEEE} Computer Society},
  year         = {2013},
  doi          = {10.1109/SP.2013.44},
}

@inproceedings{zhang2013cotscfi,
  author       = {Mingwei Zhang and
                  R. Sekar},
  title        = {Control Flow Integrity for {COTS} Binaries},
  booktitle    = {Proceedings of the 22nd {USENIX} Security Symposium ({USENIX} Security)},
  pages        = {337--352},
  publisher    = {{USENIX} Association},
  year         = {2013},
}

@inproceedings{vanderveen2015patharmor,
  author       = {Victor van der Veen and
                  Dennis Andriesse and
                  Enes G{\"{o}}ktas and
                  Ben Gras and
                  Lionel Sambuc and
                  Asia Slowinska and
                  Herbert Bos and
                  Cristiano Giuffrida},
  title        = {Practical Context-Sensitive {CFI}},
  booktitle    = {Proceedings of the 22nd {ACM} {SIGSAC} Conference on Computer and Communications Security ({CCS})},
  pages        = {927--940},
  publisher    = {{ACM}},
  year         = {2015},
  doi          = {10.1145/2810103.2813673},
}

@inproceedings{vanderveen2016typearmor,
  author       = {Victor van der Veen and
                  Enes G{\"{o}}ktas and
                  Moritz Contag and
                  Andre Pawlowski and
                  Xi Chen and
                  Sanjay Rawat and
                  Herbert Bos and
                  Thorsten Holz and
                  Elias Athanasopoulos and
                  Cristiano Giuffrida},
  title        = {A Tough Call: Mitigating Advanced Code-Reuse Attacks at the Binary
                  Level},
  booktitle    = {Proceedings of the 37th {IEEE} Symposium on Security and Privacy ({S\&P})},
  pages        = {934--953},
  publisher    = {{IEEE} Computer Society},
  year         = {2016},
  doi          = {10.1109/SP.2016.60},
}

@inproceedings{hu2018uct,
  author       = {Hong Hu and
                  Chenxiong Qian and
                  Carter Yagemann and
                  Simon Pak Ho Chung and
                  William R. Harris and
                  Taesoo Kim and
                  Wenke Lee},
  title        = {Enforcing Unique Code Target Property for Control-Flow Integrity},
  booktitle    = {Proceedings of the 25th {ACM} {SIGSAC} Conference on Computer and Communications Security ({CCS})},
  pages        = {1470--1486},
  publisher    = {{ACM}},
  year         = {2018},
  doi          = {10.1145/3243734.3243797},
}

@inproceedings{muntean2018taucfi,
  author       = {Paul Muntean and
                  Matthias Fischer and
                  Gang Tan and
                  Zhiqiang Lin and
                  Jens Grossklags and
                  Claudia Eckert},
  title        = {{\(\tau\)}CFI: Type-Assisted Control Flow Integrity for x86-64 Binaries},
  booktitle    = {Proceedings of the 21st International Symposium on Research in Attacks, Intrusions and Defenses ({RAID})},
  series       = {Lecture Notes in Computer Science},
  pages        = {423--444},
  publisher    = {Springer},
  year         = {2018},
  doi          = {10.1007/978-3-030-00470-5\_20},
}

@inproceedings{xu2019confirm,
  author       = {Xiaoyang Xu and
                  Masoud Ghaffarinia and
                  Wenhao Wang and
                  Kevin W. Hamlen and
                  Zhiqiang Lin},
  title        = {{CONFIRM:} Evaluating Compatibility and Relevance of Control-flow
                  Integrity Protections for Modern Software},
  booktitle    = {Proceedings of the 28th {USENIX} Security Symposium ({USENIX} Security)},
  pages        = {1805--1821},
  publisher    = {{USENIX} Association},
  year         = {2019},
}

@inproceedings{frassetto2022cfinsight,
  author       = {Tommaso Frassetto and
                  Patrick Jauernig and
                  David Koisser and
                  Ahmad{-}Reza Sadeghi},
  title        = {CFInsight: {A} Comprehensive Metric for {CFI} Policies},
  booktitle    = {Proceedings of the 29th Annual Network and Distributed System Security Symposium ({NDSS})},
  publisher    = {The Internet Society},
  year         = {2022},
}

@misc{clang_cfi,
  title = {{Control Flow Integrity}},
  author = {{LLVM Project}},
  year = {2026},
  howpublished = {\url{https://clang.llvm.org/docs/ControlFlowIntegrity.html}},
  note = {Accessed 2026-05-12}
}

@inproceedings{villa2019nvbit,
  author       = {Oreste Villa and
                  Mark Stephenson and
                  David W. Nellans and
                  Stephen W. Keckler},
  title        = {NVBit: {A} Dynamic Binary Instrumentation Framework for {NVIDIA} GPUs},
  booktitle    = {Proceedings of the 52nd Annual {IEEE/ACM} International Symposium on Microarchitecture ({MICRO})},
  pages        = {372--383},
  publisher    = {{ACM}},
  year         = {2019},
  doi          = {10.1145/3352460.3358307},
}

@misc{nvidia_ptx,
  title = {{Parallel Thread Execution ISA}},
  author = {{NVIDIA}},
  year = {2026},
  howpublished = {\url{https://docs.nvidia.com/cuda/parallel-thread-execution/}},
  note = {Accessed 2026-05-12}
}

@misc{nvidia_binary_utils,
  title = {{CUDA Binary Utilities}},
  author = {{NVIDIA}},
  year = {2026},
  howpublished = {\url{https://docs.nvidia.com/cuda/cuda-binary-utilities/}},
  note = {Accessed 2026-05-12}
}

@misc{nvidia_programming_guide,
  title = {{CUDA C++ Programming Guide}},
  author = {{NVIDIA}},
  year = {2026},
  howpublished = {\url{https://docs.nvidia.com/cuda/cuda-c-programming-guide/}},
  note = {Accessed 2026-05-12}
}

@misc{nvidia_nvcc,
  title = {{CUDA Compiler Driver NVCC}},
  author = {{NVIDIA}},
  year = {2026},
  howpublished = {\url{https://docs.nvidia.com/cuda/cuda-compiler-driver-nvcc/}},
  note = {Accessed 2026-05-14}
}

@inproceedings{gpushield2022,
  author       = {Jaewon Lee and
                  Yonghae Kim and
                  Jiashen Cao and
                  Euna Kim and
                  Jaekyu Lee and
                  Hyesoon Kim},
  title        = {Securing {GPU} via region-based bounds checking},
  booktitle    = {Proceedings of the 49th Annual International Symposium on Computer Architecture ({ISCA})},
  pages        = {27--41},
  publisher    = {{ACM}},
  year         = {2022},
  doi          = {10.1145/3470496.3527420},
}

@inproceedings{guardian2024,
  author       = {Manos Pavlidakis and
                  Giorgos Vasiliadis and
                  Stelios Mavridis and
                  Anargyros Argyros and
                  Antony Chazapis and
                  Angelos Bilas},
  title        = {Guardian: Safe {GPU} Sharing in Multi-Tenant Environments},
  booktitle    = {Proceedings of the 25th International Middleware Conference ({Middleware})},
  pages        = {313--326},
  publisher    = {{ACM}},
  year         = {2024},
  doi          = {10.1145/3652892.3700768},
}

@article{gpuarmor2025,
  author       = {Mohamed Tarek Ibn Ziad and
                  Sana Damani and
                  Mark Stephenson and
                  Stephen W. Keckler and
                  Aamer Jaleel},
  title        = {GPUArmor: {A} Hardware-Software Co-design for Efficient and Scalable
                  Memory Safety on GPUs},
  journal      = {CoRR},
  volume       = {abs/2502.17780},
  year         = {2025},
  doi          = {10.48550/ARXIV.2502.17780},
  eprinttype   = {arXiv},
  eprint       = {2502.17780},
}

@article{cucatch2023,
  author       = {Mohamed Tarek Ibn Ziad and
                  Sana Damani and
                  Aamer Jaleel and
                  Stephen W. Keckler and
                  Mark Stephenson},
  title        = {cuCatch: {A} Debugging Tool for Efficiently Catching Memory Safety
                  Violations in {CUDA} Applications},
  journal      = {Proc. {ACM} Program. Lang.},
  volume       = {7},
  number       = {{PLDI}},
  pages        = {124--147},
  year         = {2023},
  doi          = {10.1145/3591225},
}

@inproceedings{lu2026cusafe,
  title = {{CuSafe: Capturing Memory Corruption on NVIDIA GPUs}},
  author = {Lu, Hongyi and Zhang, Fengwei and Zhang, Zhenkai and Wang, Shuai and Guo, Yanan},
  booktitle = {Proceedings of the 35th {USENIX} Security Symposium ({USENIX} Security)},
  year = {2026},
  publisher = {USENIX Association},
  url = {https://www.usenix.org/conference/usenixsecurity26/cycle1-accepted-papers}
}

@inproceedings{mai2023honeycomb,
  author       = {Haohui Mai and
                  Jiacheng Zhao and
                  Hongren Zheng and
                  Yiyang Zhao and
                  Zibin Liu and
                  Mingyu Gao and
                  Cong Wang and
                  Huimin Cui and
                  Xiaobing Feng and
                  Christos Kozyrakis},
  title        = {Honeycomb: Secure and Efficient {GPU} Executions via Static Validation},
  booktitle    = {Proceedings of the 17th {USENIX} Symposium on Operating Systems Design and Implementation ({OSDI})},
  pages        = {155--172},
  publisher    = {{USENIX} Association},
  year         = {2023},
}

@inproceedings{ivanov2023sage,
  author       = {Andrei Ivanov and
                  Benjamin Rothenberger and
                  Arnaud Dethise and
                  Marco Canini and
                  Torsten Hoefler and
                  Adrian Perrig},
  title        = {{SAGE:} Software-based Attestation for {GPU} Execution},
  booktitle    = {Proceedings of the 2023 {USENIX} Annual Technical Conference ({USENIX} {ATC})},
  pages        = {485--499},
  publisher    = {{USENIX} Association},
  year         = {2023},
}

@misc{riscv_cfi,
  title = {{RISC-V Control-flow Integrity Extensions}},
  author = {{RISC-V International}},
  year = {2026},
  howpublished = {\url{https://docs.riscv.org/reference/isa/priv/priv-cfi.html}},
  note = {Accessed 2026-05-14}
}

@article{shoushtary2024controlflow,
  author       = {Mojtaba Abaie Shoushtary and
                  Jordi Tubella Murgadas and
                  Antonio Gonz{\'{a}}lez},
  title        = {Control Flow Management in Modern GPUs},
  journal      = {CoRR},
  volume       = {abs/2407.02944},
  year         = {2024},
  doi          = {10.48550/ARXIV.2407.02944},
  eprinttype   = {arXiv},
  eprint       = {2407.02944},
}

@article{roemer2012rop,
  author       = {Ryan Roemer and
                  Erik Buchanan and
                  Hovav Shacham and
                  Stefan Savage},
  title        = {Return-Oriented Programming: Systems, Languages, and Applications},
  journal      = {{ACM} Trans. Inf. Syst. Secur.},
  volume       = {15},
  number       = {1},
  pages        = {2:1--2:34},
  year         = {2012},
  doi          = {10.1145/2133375.2133377},
}

@inproceedings{davi2014stitching,
  author       = {Lucas Davi and
                  Ahmad{-}Reza Sadeghi and
                  Daniel Lehmann and
                  Fabian Monrose},
  title        = {Stitching the Gadgets: On the Ineffectiveness of Coarse-Grained Control-Flow
                  Integrity Protection},
  booktitle    = {Proceedings of the 23rd {USENIX} Security Symposium ({USENIX} Security)},
  pages        = {401--416},
  publisher    = {{USENIX} Association},
  year         = {2014},
}

@inproceedings{schuster2015coop,
  author       = {Felix Schuster and
                  Thomas Tendyck and
                  Christopher Liebchen and
                  Lucas Davi and
                  Ahmad{-}Reza Sadeghi and
                  Thorsten Holz},
  title        = {Counterfeit Object-oriented Programming: On the Difficulty of Preventing
                  Code Reuse Attacks in {C++} Applications},
  booktitle    = {Proceedings of the 36th {IEEE} Symposium on Security and Privacy ({S\&P})},
  pages        = {745--762},
  publisher    = {{IEEE} Computer Society},
  year         = {2015},
  doi          = {10.1109/SP.2015.51},
}

@inproceedings{evans2015controljujutsu,
  author       = {Isaac Evans and
                  Fan Long and
                  Ulziibayar Otgonbaatar and
                  Howard E. Shrobe and
                  Martin C. Rinard and
                  Hamed Okhravi and
                  Stelios Sidiroglou{-}Douskos},
  title        = {Control Jujutsu: On the Weaknesses of Fine-Grained Control Flow Integrity},
  booktitle    = {Proceedings of the 22nd {ACM} {SIGSAC} Conference on Computer and Communications Security ({CCS})},
  pages        = {901--913},
  publisher    = {{ACM}},
  year         = {2015},
  doi          = {10.1145/2810103.2813646},
}

@inproceedings{conti2015losingcontrol,
  author       = {Mauro Conti and
                  Stephen Crane and
                  Lucas Davi and
                  Michael Franz and
                  Per Larsen and
                  Marco Negro and
                  Christopher Liebchen and
                  Mohaned Qunaibit and
                  Ahmad{-}Reza Sadeghi},
  title        = {Losing Control: On the Effectiveness of Control-Flow Integrity under
                  Stack Attacks},
  booktitle    = {Proceedings of the 22nd {ACM} {SIGSAC} Conference on Computer and Communications Security ({CCS})},
  pages        = {952--963},
  publisher    = {{ACM}},
  year         = {2015},
  doi          = {10.1145/2810103.2813671},
}

@article{burow2017cfisurvey,
  author       = {Nathan Burow and
                  Scott A. Carr and
                  Joseph Nash and
                  Per Larsen and
                  Michael Franz and
                  Stefan Brunthaler and
                  Mathias Payer},
  title        = {Control-Flow Integrity: Precision, Security, and Performance},
  journal      = {{ACM} Comput. Surv.},
  volume       = {50},
  number       = {1},
  pages        = {16:1--16:33},
  year         = {2017},
  doi          = {10.1145/3054924},
}

@misc{intel_cet,
  title = {{A Technical Look at Intel Control-Flow Enforcement Technology}},
  author = {{Intel}},
  year = {2020},
  howpublished = {\url{https://www.intel.com/content/www/us/en/developer/articles/technical/technical-look-control-flow-enforcement-technology.html}},
  note = {Accessed 2026-05-14}
}

@misc{microsoft_cfg,
  title = {{/guard: Enable Control Flow Guard}},
  author = {{Microsoft}},
  year = {2025},
  howpublished = {\url{https://learn.microsoft.com/cpp/build/reference/guard-enable-control-flow-guard}},
  note = {Accessed 2026-05-14}
}

@misc{arm_pac,
  title = {{Improving Control Flow Integrity with Pointer Authentication}},
  author = {{Apple}},
  year = {2026},
  howpublished = {\url{https://developer.apple.com/documentation/apple-silicon/improving-control-flow-integrity-with-pointer-authentication}},
  note = {Accessed 2026-05-14}
}

@misc{clang_shadowcallstack,
  title = {{ShadowCallStack}},
  author = {{LLVM Project}},
  year = {2026},
  howpublished = {\url{https://clang.llvm.org/docs/ShadowCallStack.html}},
  note = {Accessed 2026-05-14}
}

@inproceedings{di2016overflow,
  author       = {Bang Di and
                  Jianhua Sun and
                  Hao Chen},
  title        = {A Study of Overflow Vulnerabilities on GPUs},
  booktitle    = {Proceedings of the 13th {IFIP} {WG} 10.3 International Conference on Network and Parallel Computing ({NPC})},
  series       = {Lecture Notes in Computer Science},
  pages        = {103--115},
  year         = {2016},
  doi          = {10.1007/978-3-319-47099-3\_9},
}

@article{di2015cudabuffer,
  author       = {Andrea Miele},
  title        = {Buffer overflow vulnerabilities in {CUDA:} a preliminary analysis},
  journal      = {J. Comput. Virol. Hacking Tech.},
  volume       = {12},
  number       = {2},
  pages        = {113--120},
  year         = {2016},
  doi          = {10.1007/S11416-015-0251-1},
}

@inproceedings{zhu2026gpuaslr,
  title = {{Demystifying and Exploiting ASLR on NVIDIA GPUs}},
  author = {Zhu, Ruofan and Chen, Ganhao and Shen, Wenbo and Zhang, Lyuye and Shen, Dakun and Chang, Rui and Guo, Yanan},
  booktitle = {Proceedings of the 47th {IEEE} Symposium on Security and Privacy ({S\&P})},
  year = {2026},
  publisher = {IEEE},
}

@inproceedings{roh2026ghost,
  title = {{GHost in the SHELL: A GPU-to-Host Memory Attack and Its Mitigation}},
  author = {Roh, Sihyun and Choi, Woohyuk and Chung, Jaeyoung and Lee, Yoochan and Song, Suhwan and Lee, Byoungyoung},
  booktitle = {Proceedings of the 47th {IEEE} Symposium on Security and Privacy ({S\&P})},
  year = {2026},
  publisher = {IEEE},
}

@article{dipietro2016cudaleaks,
  author       = {Roberto Di Pietro and
                  Flavio Lombardi and
                  Antonio Villani},
  title        = {{CUDA} Leaks: {A} Detailed Hack for {CUDA} and a (Partial) Fix},
  journal      = {{ACM} Trans. Embed. Comput. Syst.},
  volume       = {15},
  number       = {1},
  pages        = {15:1--15:25},
  year         = {2016},
  doi          = {10.1145/2801153},
}

@misc{nvidia_compute_sanitizer,
  title = {{Compute Sanitizer}},
  author = {{NVIDIA}},
  year = {2026},
  howpublished = {\url{https://docs.nvidia.com/cuda/compute-sanitizer/}},
  note = {Accessed 2026-05-14}
}

@inproceedings{stephenson2015sassi,
  author       = {Mark Stephenson and
                  Siva Kumar Sastry Hari and
                  Yunsup Lee and
                  Eiman Ebrahimi and
                  Daniel R. Johnson and
                  David W. Nellans and
                  Mike O'Connor and
                  Stephen W. Keckler},
  title        = {Flexible software profiling of {GPU} architectures},
  booktitle    = {Proceedings of the 42nd Annual International Symposium on Computer Architecture ({ISCA})},
  pages        = {185--197},
  publisher    = {{ACM}},
  year         = {2015},
  doi          = {10.1145/2749469.2750375},
}

@inproceedings{tsai2021nvbitfi,
  author       = {Timothy Tsai and
                  Siva Kumar Sastry Hari and
                  Michael B. Sullivan and
                  Oreste Villa and
                  Stephen W. Keckler},
  title        = {NVBitFI: Dynamic Fault Injection for GPUs},
  booktitle    = {Proceedings of the 51st Annual {IEEE/IFIP} International Conference on Dependable Systems and Networks ({DSN})},
  pages        = {284--291},
  publisher    = {{IEEE}},
  year         = {2021},
  doi          = {10.1109/DSN48987.2021.00041},
}

@inproceedings{diamos2010ocelot,
  author       = {Gregory Frederick Diamos and
                  Andrew Kerr and
                  Sudhakar Yalamanchili and
                  Nathan Clark},
  title        = {Ocelot: a dynamic optimization framework for bulk-synchronous applications
                  in heterogeneous systems},
  booktitle    = {Proceedings of the 19th International Conference on Parallel Architectures and Compilation Techniques ({PACT})},
  pages        = {353--364},
  publisher    = {{ACM}},
  year         = {2010},
  doi          = {10.1145/1854273.1854318},
}

@inproceedings{wan2026nvlift,
  title = {{NVLift: Lifting NVIDIA GPU Assembly to LLVM IR for Downstream Security Applications}},
  author = {Wan, Junpeng and Tan, Louis Zheng-Hua and Tian, Dave (Jing)},
  booktitle = {Proceedings of the Workshop on Binary Analysis Research ({BAR})},
  year = {2026},
  publisher = {Internet Society},
  doi = {10.14722/bar.2026.23028}
}

@inproceedings{volos2018graviton,
  author       = {Stavros Volos and
                  Kapil Vaswani and
                  Rodrigo Bruno},
  title        = {Graviton: Trusted Execution Environments on GPUs},
  booktitle    = {Proceedings of the 13th {USENIX} Symposium on Operating Systems Design and Implementation ({OSDI})},
  pages        = {681--696},
  publisher    = {{USENIX} Association},
  year         = {2018},
}

@inproceedings{hunt2020telekine,
  author       = {Tyler Hunt and
                  Zhipeng Jia and
                  Vance Miller and
                  Ariel Szekely and
                  Yige Hu and
                  Christopher J. Rossbach and
                  Emmett Witchel},
  title        = {Telekine: Secure Computing with Cloud GPUs},
  booktitle    = {Proceedings of the 17th {USENIX} Symposium on Networked Systems Design and Implementation ({NSDI})},
  pages        = {817--833},
  publisher    = {{USENIX} Association},
  year         = {2020},
}

@inproceedings{jang2019hix,
  author       = {Insu Jang and
                  Adrian Tang and
                  Taehoon Kim and
                  Simha Sethumadhavan and
                  Jaehyuk Huh},
  title        = {Heterogeneous Isolated Execution for Commodity GPUs},
  booktitle    = {Proceedings of the 24th {ACM} International Conference on Architectural Support for Programming Languages and Operating Systems ({ASPLOS})},
  pages        = {455--468},
  publisher    = {{ACM}},
  year         = {2019},
  doi          = {10.1145/3297858.3304021},
}

@misc{nvidia_mig,
  title = {{NVIDIA Multi-Instance GPU User Guide}},
  author = {{NVIDIA}},
  year = {2026},
  howpublished = {\url{https://docs.nvidia.com/datacenter/tesla/mig-user-guide/}},
  note = {Accessed 2026-05-14}
}

@misc{nvidia_confidential_computing,
  title = {{NVIDIA Confidential Computing}},
  author = {{NVIDIA}},
  year = {2026},
  howpublished = {\url{https://www.nvidia.com/en-us/data-center/solutions/confidential-computing/}},
  note = {Accessed 2026-05-14}
}

@misc{repo_cuda_quantum,
  title = {{CUDA-Q}},
  author = {{NVIDIA}},
  year = {2026},
  howpublished = {\url{https://github.com/NVIDIA/cuda-quantum}},
  note = {GitHub repository; accessed 2026-05-23}
}

@misc{repo_goofit,
  title = {{GooFit}},
  author = {{GooFit Contributors}},
  year = {2026},
  howpublished = {\url{https://github.com/GooFit/GooFit}},
  note = {GitHub repository; accessed 2026-05-23}
}

@misc{repo_kokkos,
  title = {{Kokkos}},
  author = {{Kokkos Contributors}},
  year = {2026},
  howpublished = {\url{https://github.com/kokkos/kokkos}},
  note = {GitHub repository; accessed 2026-05-23}
}

@misc{repo_cuda_samples,
  title = {{CUDA Samples}},
  author = {{NVIDIA}},
  year = {2026},
  howpublished = {\url{https://github.com/NVIDIA/cuda-samples}},
  note = {GitHub repository; accessed 2026-05-23}
}

@misc{repo_rawspec,
  title = {{rawspec}},
  author = {{UCBerkeleySETI}},
  year = {2026},
  howpublished = {\url{https://github.com/UCBerkeleySETI/rawspec}},
  note = {GitHub repository; accessed 2026-05-23}
}

@misc{repo_ptypy,
  title = {{ptypy}},
  author = {{ptypy Contributors}},
  year = {2026},
  howpublished = {\url{https://github.com/ptycho/ptypy}},
  note = {GitHub repository; accessed 2026-05-23}
}

@misc{repo_empi,
  title = {{empi}},
  author = {{empi Contributors}},
  year = {2026},
  howpublished = {\url{https://github.com/develancer/empi}},
  note = {GitHub repository; accessed 2026-05-23}
}

@misc{repo_drjit,
  title = {{Dr.Jit}},
  author = {{Dr.Jit Contributors}},
  year = {2026},
  howpublished = {\url{https://github.com/mitsuba-renderer/drjit}},
  note = {GitHub repository; accessed 2026-05-23}
}

@misc{repo_ppl_cuda_smc,
  title = {{PPL-CUDA-SMC}},
  author = {{PPL-CUDA-SMC Contributors}},
  year = {2026},
  howpublished = {\url{https://github.com/JoeyOhman/PPL-CUDA-SMC}},
  note = {GitHub repository; accessed 2026-05-23}
}

@misc{goofit_issue242,
  title = {{GooFit Issue \#242: kMatrix/Amp3Body}},
  author = {{GooFit Contributors}},
  year = {2026},
  howpublished = {\url{https://github.com/GooFit/GooFit/issues/242}},
  note = {GitHub issue; accessed 2026-05-23}
}

@misc{repo_sv_sim,
  title = {{SV-Sim}},
  author = {{Pacific Northwest National Laboratory}},
  year = {2026},
  howpublished = {\url{https://github.com/pnnl/SV-Sim}},
  note = {GitHub repository; accessed 2026-05-23}
}

@misc{repo_dm_sim,
  title = {{DM-Sim}},
  author = {{Pacific Northwest National Laboratory}},
  year = {2026},
  howpublished = {\url{https://github.com/pnnl/DM-Sim}},
  note = {GitHub repository; accessed 2026-05-23}
}

\appendices
%\setcounter{table}{0}
%\setcounter{figure}{0}
%\renewcommand{\thetable}{A.\arabic{table}}
%\renewcommand{\thefigure}{A.\arabic{figure}}
%\makeatletter
%\renewcommand{\theHtable}{appendix.table.\arabic{table}}
%\renewcommand{\theHfigure}{appendix.figure.\arabic{figure}}
%\makeatother
\section{Open Science}
\label{app:open-science}

The anonymized \tool{} tool and reproduction artifact are available at
\url{https://anonymous.4open.science/r/warpguard-anon/README.md}. The package
includes the tool source, CUDA/SASS fixtures, reproduction scripts, public-code
recipes, expected outputs, and runbooks for the evidence tiers reported in the
paper.

The artifact is organized as a single codebase with two entry points. The
\texttt{tool/} tree contains the \tool{} implementation and developer-facing
CLI. The \texttt{artifact/} tree contains paper-facing reproduction wrappers
that call the same tool code. This avoids a separate artifact implementation:
the scripts used to reproduce the paper exercise the same source tree exposed
to readers.
% Tables~\ref{tab:app-evidence-reader}
% --\ref{tab:app-impl-support} define terms, trust boundary, SASS classes,
% runtime profiles, and the support boundary. Tables~\ref{tab:app-rewrite-forms}
% --\ref{tab:app-security-evidence} summarize rewrite forms, corpus lanes,
% residual surface, and security evidence. Table~\ref{tab:app-backend-evidence}
% reports backend-cost and callback-free evidence under separate denominators.

\section{Supplementary Evidence}
\label{app:supplementary-data}

This appendix lists the audit evidence and supporting tables; generated result
files contain the full case-level matrices. The tables below define the compact
terms used in the main text, give the supporting trust and implementation
boundaries, summarize the corpus and attack evidence, and record backend-cost
and callback-free evidence under separate denominators.

\setcounter{table}{0}
\renewcommand{\thetable}{A\arabic{table}}
\makeatletter
\renewcommand{\theHtable}{appendix.table.\arabic{table}}
\makeatother

\begingroup
\scriptsize
\setlength{\tabcolsep}{2.2pt}
\renewcommand{\arraystretch}{0.86}
\setlength{\floatsep}{4pt plus 1pt minus 1pt}
\setlength{\textfloatsep}{4pt plus 1pt minus 1pt}
\setlength{\intextsep}{3pt plus 1pt minus 1pt}
\setlength{\dblfloatsep}{4pt plus 1pt minus 1pt}
\setlength{\dbltextfloatsep}{4pt plus 1pt minus 1pt}
\setlength{\abovecaptionskip}{2pt}
\setlength{\belowcaptionskip}{0pt}
\setcounter{topnumber}{5}
\setcounter{bottomnumber}{4}
\setcounter{totalnumber}{8}
\setcounter{dbltopnumber}{3}
\renewcommand{\topfraction}{0.95}
\renewcommand{\bottomfraction}{0.85}
\renewcommand{\textfraction}{0.05}
\renewcommand{\floatpagefraction}{0.70}
\renewcommand{\dbltopfraction}{0.95}
\renewcommand{\dblfloatpagefraction}{0.70}

\begin{table}[!htbp]
\centering
\caption{Representative SASS control-flow classes.}
\label{tab:app-sass-classes}
\resizebox{\columnwidth}{!}{%
\begin{tabular}{@{}p{0.22\columnwidth}p{0.20\columnwidth}p{0.46\columnwidth}@{}}
\toprule
Class & Examples & Security complication \\
\midrule
call/return & \texttt{CALL}, \texttt{RET} & Continuation state may be
register-backed, spilled, ABI-specific, or optimized away. \\
predicated branch & \texttt{@P BRA} & Enabled lanes, not only the static edge,
determine which dynamic transfer executed. \\
indirect transfer & \texttt{BRX}, \texttt{JMX} & The target set may be
unrecoverable in closed or relocation-poor binaries. \\
termination & \texttt{EXIT}, trap & Invalid transfers must fail closed rather
than being released as ordinary control flow. \\
SIMT state & \texttt{BSSY}, barrier/sync & Reconvergence state is classified
separately and is not treated as a normal CFI target edge. \\
\bottomrule
\end{tabular}
}%
\end{table}

\begin{table}[!htbp]
\centering
\caption{Runtime profiles.}
\label{tab:app-runtime-profiles}
\resizebox{\columnwidth}{!}{%
\begin{tabular}{@{}p{0.30\columnwidth}p{0.58\columnwidth}@{}}
\toprule
Profile & Scope \\
\midrule
diagnostic & Counters/traces only; no prevention claim. \\
detect-only & Records violations and continues. \\
backward-only & Authenticated returns only. \\
forward-only & Selective forward checks only. \\
full enforcement & Backward plus recoverable forward sites; fail closed. \\
low-telemetry enforcement & Same checked transfers with less reporting. \\
coverage-reduced & Skips kernels, sites, counters, or a CFI direction; separate
denominator. \\
attribution-only & Helper-cost measurement; not protected. \\
\bottomrule
\end{tabular}
}%
\end{table}

\newpage

\begin{table}[!htbp]
\centering
\caption{Trusted-state boundary.}
\label{tab:app-trust-boundary}
\resizebox{\columnwidth}{!}{%
\begin{tabular}{@{}p{0.24\columnwidth}p{0.18\columnwidth}p{0.48\columnwidth}@{}}
\toprule
State & Attacker access & Role \\
\midrule
Application global, local, shared, and heap memory & Read/write if reachable by
a device bug & Attack input. May hold function pointers, dispatch tables,
jump-table data, continuation slots, or writable metadata. \\
Telemetry counters and violation records & Writable & Diagnostic
only; never authorizes a transfer. \\
Writable shadow and forward records & Writable & Checked or
authenticated before use. \\
WG-NVBit private state and keys & Trusted in this model & Authorizes broad
metadata-authenticated CFI cases. \\
WG-PC patch-cache private state & Trusted in this model & Selects verified
policy-bound patch plans for reviewed \texttt{sm\_89} surfaces. \\
\bottomrule
\end{tabular}
}%
\end{table}

\begin{table}[!htbp]
\centering
\caption{Callback-free SASS rewrite forms.}
\label{tab:app-rewrite-forms}
\resizebox{\columnwidth}{!}{%
\begin{tabular}{@{}p{0.25\columnwidth}p{0.25\columnwidth}p{0.40\columnwidth}@{}}
\toprule
Rewrite form & Lane and edge & Required condition / failure mode \\
\midrule
Fixed-size checked patch & WG-PC; simple transfer & Fits without live
state clobber; otherwise reject/fail closed. \\
Trampoline patch & WG-ST/WG-PC; larger check path & Safe branch window and
verified target; otherwise reject/fail closed. \\
Callee clone & WG-PC; call/return copy & Preserves register, predicate,
and return behavior; otherwise unsupported. \\
Scoped dynamic shadow & WG-PC; runtime shadow checks & Reviewed
\texttt{sm\_89} manifest and policy-bound cache entry; otherwise fail closed. \\
\bottomrule
\end{tabular}
}%
\end{table}

\begin{table*}[!t]
\centering
\caption{Evidence terms and exposure categories. Denominators are separate
measurement units; attack levels describe realism, not severity.}
\label{tab:app-evidence-reader}
\scriptsize
\setlength{\tabcolsep}{3pt}
\begin{tabular}{@{}p{0.18\textwidth}p{0.38\textwidth}p{0.36\textwidth}@{}}
\toprule
Term & Meaning & Use in the paper \\
\midrule
\multicolumn{3}{@{}l}{\textit{Evaluation denominators}}\\
Raw SASS sites & Instruction-level recovered control-flow sites in executed
SASS. & Recovery scale; not a protected-coverage numerator. \\
Policy entries & Site or site-group outcomes after policy generation. & Accounting
for protected, fixed-edge, unsupported, profile-excluded, fallback, and
no-surface outcomes. \\
Forward target-set entries & Per-site records for recoverable indirect
targets. & Forward support and precision; provenance entries use a finer audit
granularity. \\
Protected-active executions & Dynamic executions whose selected profile keeps active
checks. & Runtime evidence for checked transfers under that profile. \\
Attack cases & Regression cases: native, detect-only, enforce, benign,
boundary, or tamper. & Security outcomes; evidence classes can overlap with
the protected attack/benign denominator. \\
Backend evidence & WG-NVBit, WG-ST, and WG-PC evidence lanes. & Cost, matched
timing, and static prevention are compared only within their denominators. \\
\addlinespace[2pt]
\multicolumn{3}{@{}l}{\textit{Return-exposure categories}}\\
Return-free & No dynamic return is present. & Return-free code has no backward CFI
consumption site. \\
Register-backed & A return exists, but continuation state remains register-backed. & A
\texttt{RET} count alone does not imply writable continuation exposure. \\
Frame-only & Stack or frame state exists, but not as a useful writable continuation
consumed by \texttt{RET}. & Frame activity is separated from exploitable return
state. \\
Static candidate & Static SASS shows a writable spill/reload continuation candidate. & Static
candidate; still missing confirmed attacker reachability and diversion. \\
Confirmed diversion & Native execution confirms diversion through attacker-controlled writable
continuation state. & Confirmed backward hijack evidence. \\
\addlinespace[2pt]
\multicolumn{3}{@{}l}{\textit{Attack realism ladder}}\\
L0 & Benign protected returns and forward transfers. & Valid checked paths run
without false violations in the tested cases. \\
L1 & Isolated return hijack, function-pointer overwrite, token tamper, or
metadata tamper. & Single corrupted control object reaches a protected check. \\
L2 & Useful-write gadget, two-return mini-chain, or divergent-lane return. &
Composed code-reuse shapes stop at the first invalid protected transfer. \\
L3 & ABI-stress continuation exposure; seeded image-pipeline, benchmark,
CUDA-sample-style, or ML-shaped dispatch case. & Source-assisted or
evaluation-seeded cases inside realistic compiler/benchmark/application
structure. \\
L4 & Guo-style/adapted Guo cases \cite{guo2024gpu} and validated public-code
consumption cases. & Prior or public-code evidence recovered and checked at
SASS consumption. \\
\bottomrule
\end{tabular}
\end{table*}

\begin{table*}[!t]
\centering
\caption{Runtime facts and support boundary.}
\label{tab:app-impl-support}
\scriptsize
\setlength{\tabcolsep}{3pt}
\begin{tabular}{@{}p{0.18\textwidth}p{0.28\textwidth}p{0.46\textwidth}@{}}
\toprule
Site form & Runtime action & Boundary \\
\midrule
Direct \texttt{CALL} & Push fall-through continuation. & Fixed call target under
code integrity. \\
Indirect \texttt{CALL} & Check target, then push continuation. & Unsupported if
target or continuation evidence is insufficient. \\
\texttt{RET}/\texttt{RET.ABS} & Check consumed return operand. & Consumption-time
check; earlier spills are untrusted. \\
Reloaded return & Check the reloaded return register. & Spill/reload remains a
static candidate until write reachability and diversion are confirmed. \\
Indirect branch & Check \texttt{BRX}, \texttt{BRXU}, \texttt{JMX}, or \texttt{JMPX}
when target materialization and target-set recovery succeed. & Unsupported when
target evidence is opaque. \\
Predication and SIMT & Apply guards to active lanes. & Reconvergence/barriers are
not ordinary CFI target edges. \\
\bottomrule
\end{tabular}
\end{table*}

\begin{table*}[!t]
\centering
\caption{Evaluation corpus lanes. Counts are non-additive.}
\label{tab:app-corpus-lanes}
\scriptsize
\setlength{\tabcolsep}{3pt}
\begin{tabular}{@{}p{0.21\textwidth}r r r p{0.43\textwidth}@{}}
\toprule
Lane & Artifacts & Returns & Fwd. & Use and examples \\
\midrule
Controlled\slash generated & 9 & 62 & 25/0 & Return hijack, useful-write,
mini-chain, function-pointer overwrite. \\
CUPTI\slash source utilities & 12 & 0 & 2/0 & Classification and compatibility
breadth. \\
Standard\slash source roots & 35 & 328 & 127/0 & CUDA Samples, CUTLASS, Rodinia,
Parboil, PolyBench-GPU. \\
CUDA toolkit library & 1 & 934 & 0/0 & Bounded \texttt{libcudadevrt.a}
sampling for runtime/library SASS surface. \\
External binary samples & 20 & 19 & 0/6 & Binary-only and extension samples. \\
Triton cubin sweep & 53 & 0 & 0/6 & JIT cubins for return-free and unsupported
indirect cases. \\
\bottomrule
\end{tabular}
\end{table*}

\begin{table*}[!t]
\centering
\caption{Recovery, accounting, and residual surface.}
\label{tab:app-sass-residuals}
\scriptsize
\setlength{\tabcolsep}{3pt}
\begin{tabular}{@{}p{0.22\textwidth}p{0.29\textwidth}p{0.41\textwidth}@{}}
\toprule
View & Result & Meaning \\
\midrule
SASS surface & 51,621 sites; 1,343 returns; 9,727 return gadgets. & Executed
SASS denominator. \\
SASS-boundary audit & 125 artifact classifications: 86 no-surface, 22 backward-only, 4 forward-only,
13 full-CFI. & Source-visible structure is insufficient. \\
Function audit & 836 functions: 104 no-device-return, 292 callsite-only, 440
checked-return. & Function names do not define return exposure. \\
Forward recovery & 154 supported target-set entries; 15 unsupported; set size
median/p90/max 1/3/10. & Unknown target evidence stays unsupported. \\
Forward residual & 118 singleton entries; 45 multi-target entries; 94 same-site
alternatives. & Same-site semantic bending remains. \\
Symbol residual screen & 0 effectful entries/alternatives by write, memory, or
termination name heuristic. & Triage only, not a proof. \\
Backward search & 52 ABI probe/config cases and corpus triage; no local natural
confirmed memory-backed diversion. & Return/gadget counts are surface metrics. \\
Remote packs & RTX 2080 Ti (\texttt{sm\_75}), A10 (\texttt{sm\_86}), H100
80GB (\texttt{sm\_90}): 24/24 attack-smoke, 208/208 layout, 36/36 atlas. &
Portability checks, not full reruns. \\
\bottomrule
\end{tabular}
\end{table*}

\begin{table*}[!t]
\centering
\caption{Attack and metadata evidence. Native reaches attacker behavior;
detect-only records; enforce fails closed.}
\label{tab:app-security-evidence}
\scriptsize
\setlength{\tabcolsep}{3pt}
\begin{tabular}{@{}p{0.21\textwidth}p{0.35\textwidth}p{0.36\textwidth}@{}}
\toprule
Evidence class & Cases/outcome & Interpretation \\
\midrule
\multicolumn{3}{@{}l}{\textit{Controlled attack and metadata cases}}\\
Benign protected transfers & 24 compatibility cases; no false violation. & Valid
checked paths execute. \\
Backward attacks & 20 cases, 14 protected; 7 detect/enforce pairs fail closed
with \texttt{ret\_viol}. & Corrupted continuations stop before \texttt{RET}
release. \\
Forward attacks & 23 cases, 15 protected; 7 detect/enforce pairs fail closed
with \texttt{fwd\_viol}. & Function-pointer, dispatch-table, wrong-site, and
forbidden-target cases are blocked. \\
Metadata/policy tamper & 10 protected cases; 5 detect/enforce pairs fail closed.
& Writable metadata does not authorize. \\
Writable shadow baseline & Plain writable-shadow equality can be forged;
\tool{} keyed records fail closed. & Authentication is needed beyond writable
shadow state. \\
Prior-work-shaped case & Guo-style primitive and upstream-derived adapted
harness on local \texttt{sm\_89}/CUDA-13. & Checks published-shape corruption at
SASS consumption. \\
\bottomrule
\end{tabular}
\end{table*}

\begin{table*}[!t]
\centering
\caption{External public-code evidence. Cases are included only when public CUDA
code has local SASS recovery, a concrete CFI consumption site, and a validation
case. Source-only replay candidates and invalid-PC symptoms remain in the
artifact triage log until they pass this threshold. Scan counts are
protected/unsupported/direct.}
\label{tab:app-public-code-evidence}
\scriptsize
\setlength{\tabcolsep}{3pt}
\begin{tabular}{@{}p{0.16\textwidth}p{0.55\textwidth}p{0.21\textwidth}@{}}
\toprule
Class & Validation and result & Meaning \\
\midrule
Runtime tables & CUDA-Q~\cite{repo_cuda_quantum},
GooFit~\cite{repo_goofit}, and Kokkos~\cite{repo_kokkos}
post-setup device-visible
targets are sealed. Benign cases stay clean; controlled wrong-target cases raise
\texttt{fwd\_viol} and enforcement stops unauthorized results. &
Post-initialization sealing is checkable. \\
cuFFT callbacks & CUDA Samples~\cite{repo_cuda_samples},
\texttt{rawspec}~\cite{repo_rawspec}, \texttt{ptypy}~\cite{repo_ptypy},
and \texttt{empi}~\cite{repo_empi} linked callback consumers scan as
18/0/114, 26/0/145, 97/16/305, and 21/0/133. A sealed \texttt{ptypy}
wrong-callback case records 64
\texttt{fwd\_viol} checks and exits 139. & Callback consumers are supported;
intent narrowing is separate. \\
Generated callables & Dr.Jit~\cite{repo_drjit} \texttt{dr.switch()} exposes recovered
\texttt{callables[]} metadata; cubin scans 4/0/5. Same-site selectors remain
legal; a one-past-table id gives native invalid table read and WG-PC
\texttt{fwd\_viol=1}. & Generated tables are checkable; same-site residual
remains. \\
SMC table & PPL-CUDA-SMC~\cite{repo_ppl_cuda_smc} \texttt{execFuncs}:
corrupting a device
\texttt{pplFunc\_t} table to an out-of-set marker gives scan 3/0/10; native
corrupt reaches \texttt{marker=0x51c0}; detect records \texttt{fwd\_viol=1};
enforce exits 139 before marker output. & ELF data-object evidence gives a
closed target set. \\
Issue replay & GooFit issue~\#242~\cite{goofit_issue242}
kMatrix/Amp3Body replay: native aborts; detect records 414 and 690
\texttt{ret\_viol}; enforce exits 139; matched
non-kMatrix baseline has zero \texttt{ret\_viol}. & Public code reaches
protected return consumption. \\
Gate-pointer replay & PNNL SV-Sim~\cite{repo_sv_sim}/DM-Sim~\cite{repo_dm_sim}
uploaded \texttt{Gate::op} pointers
corrupted to an out-of-set marker. SV scans 42/0/7483 and DM scans 45/0/7533;
native markers are 23/46; detect records forward violations; enforce exits 139.
& Scalar gate-pointer ELF evidence recovers the target family. \\
% \addlinespace[2pt]
% \multicolumn{3}{@{}p{0.92\textwidth}@{}}{\emph{Provenance.}
% The replay index in the artifact points to pinned
% recipes for these cases; citations identify upstream GitHub projects or issue
% pages.} \\
\bottomrule
\end{tabular}
\end{table*}

\begin{table*}[!t]
\centering
\caption{Backend evidence. Only matched timing directly compares Native,
WG-NVBit, WG-ST, and WG-PC.}
\label{tab:app-backend-evidence}
\scriptsize
\setlength{\tabcolsep}{3pt}
\begin{tabular}{@{}p{0.22\textwidth}p{0.34\textwidth}p{0.36\textwidth}@{}}
\toprule
Question & Result & Claim allowed \\
\midrule
WG-NVBit reference cost & Compact epoch: 799.7\% overhead on workloads
$\geq$50 ms; compact enforcement: 927.6\%. & Broad reference backend cost. \\
WG-NVBit attribution & Callback/helper plumbing dominates; metadata, token, and
fail-closed each add below 1.3\% after plumbing. & Policy logic is not dominant. \\
Matched timing & Same two \texttt{sm\_89} timing cases, repeat 30 and warmup 5:
WG-NVBit 99.27$\times$ native, WG-ST 0.948$\times$ native, and WG-PC
0.269$\times$ native. &
Callback-free lanes remove the NVBit callback floor on the matched surface. \\
WG-PC fixture timing & Native/WG-PC medians: padded forward 0.578/0.127 ms;
relocated returns 0.614/0.164--0.166 ms; relocated target 0.584/0.134 ms;
live-GPR indirect call 0.601/0.138 ms. & Supported-manifest timing for reviewed
WG-PC fixtures; separate from WG-ST matched timing and WG-NVBit reference cost. \\
WG-PC prevention packet & 10 prevention manifests; 59 patchable release sites;
0 unsupported release sites; negative cases fail closed. & Reviewed
patch-cache manifests execute supported callback-free prevention. \\
WG-PC real-cubin breadth & 17/59 selected real cubins contain accepted
protected sites; 374 accepted sites total; unsupported plans are refused. A
3-case benign-output packet matches checksums over 31 sites and one corrupted
Function\-Pointers case fails closed. & Real cubins contain WG-PC-supported sites;
selected patched cubins preserve output. \\
WG-ST role & 59/59 selected cubins validate for scan/plan compatibility, but
WG-ST accepts 0 protected sites there. & WG-ST remains the equal-surface timing
lane. \\
\bottomrule
\end{tabular}
\end{table*}

\endgroup

\end{document}